# Geometry of the canonical Van Vleck transformation

Flemming Jørgensen

Flinten 10, 4700 Næstved, Denmark.

**Graphical table of content**

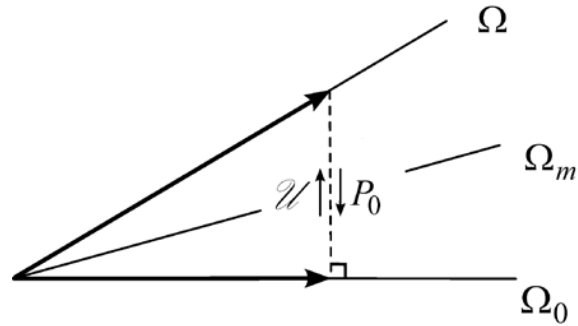

Bloch's transformation $\mathscr{U}:\Omega_0 \to \Omega$ from the zeroth order eigenspace for a perturbation problem to that of the exact eigenfunctions is simply the inverse of the projection $P_0$ on $\Omega_0$. Using a new commutation relation, the canonical Van Vleck transformation is studied as the reflection in the bisecting plane $\Omega_m$ and as a rotation around $\Omega_0 \cap \Omega$. Klein's unique characteristic is derived from simple geometry, seen as a direct consequence of Jørgensen's and extended.

**Abstract**

A Van Vleck transformation $U = e^g$; $g = -g^\dagger$, to an effective Hamiltonian changes an orthonormal basis in the zeroth order eigenspace $\Omega_0$ to one in the subspace $\Omega$ of the corresponding exact eigenvectors. The *canonical* $U_c = e^{g_c}$ is the only where $g$ is *odd*. Jørgensen's, theorem of uniqueness reveals that $U_c$ equals $U_{c'} \equiv P(P_0 P P_0)^{-\frac{1}{2}} + Q(Q_0 Q Q_0)^{-\frac{1}{2}}$ where $P_0/P$ project on $\Omega_0/\Omega$ and $Q_0/Q$ on $\Omega_0^\perp/\Omega^\perp$. By Klein's theorem of uniqueness, $u_c = P(P_0 P P_0)^{-\frac{1}{2}}$ is the mapping $u:\Omega_0 \to \Omega$ which changes an orthonormal basis in $\Omega_0$ minimally.

In the present paper, Klein's theorem is developed, proven by simple geometry and also as a direct consequence of Jørgensen's. It is shown that $U_{c'}$ equals $|S|^{-1}S = S|S|^{-1}$ where $S \equiv PP_0 + QQ_0$ satisfies $SS^\dagger = S^\dagger S \equiv |S|^2$. These commutations simplify earlier proofs, lead to $g_c$ in terms of $P_0$ and $P$ and to a series of geometrical interpretations, all easily illustrated in the elementary case where $\Omega_0$ and $\Omega$ are 2-dimensional planes in the 3-dimensional space. Thus $U_c : \Omega_0 \to \Omega$ is the reflection in the plane $\Omega_m$ between $\Omega_0$ and $\Omega$ as well as a rotation around their line of intersection.

**1: Introduction**

Let us briefly recapitulate the basic ideas behind the construction and use of an effective Hamiltonian. Consider a perturbation problem

$$H = H_0 + \lambda V \tag{1.1}$$

where $H_0$ is the zeroth order Hamiltonian and $\lambda \in [0,1]$ is an interpolation parameter by which the perturbation $V$ can be turned on continuously. We wish to determine those eigenvalues of $H$ which arise from a certain zeroth order eigenvalue (sometimes from several lying close together). Let $P_0$ be the projector on the zeroth order eigenspace $\Omega_0$, let $P$ be the projector on the subspace $\Omega$ of the corresponding exact eigenvectors and let $Q_0 = 1 - P_0$ and $Q = 1 - P$ be the projectors on the orthogonal complements $\Omega_0^\perp$ and $\Omega^\perp$. In the theory of transformation to an effective Hamiltonian as considered in the present paper, four different sets of transformation operators are fundamental: The set $\mathscr{B}_{INV}(\Omega_0 \to \Omega)$ consists all invertible operators $U$ which satisfy the condition

$$UP_0 U^{-1} = P \quad (\Leftrightarrow UP_0 = PU), \tag{1.2}$$

and $\mathscr{B}_{UNI}(\Omega_0 \to \Omega)$ is the subset of the unitary operators. The set $\mathscr{B}_{inv}(\Omega_0 \to \Omega)$ consists of all operators of the form $u = PuP_0$ for which the mapping $u : \Omega_0 \to \Omega$ has an inverse $t : \Omega \to \Omega_0$ of the form $t = P_0 tP$ so that

$$tu = P_0 \quad \text{and} \quad ut = P. \tag{1.3}$$

And $\mathscr{B}_{uni}(\Omega_0 \to \Omega)$ is the subset of those $u$ which satisfy $t = u^\dagger$ so that the inner product is conserved in the mapping $u : \Omega_0 \to \Omega$ and its inverse $u^\dagger : \Omega \to \Omega_0$. Defining sets like $\mathscr{B}_{INV}(\Omega \to \Omega_0)$ and $\mathscr{B}_{UNI}(\Omega_0^\perp \to \Omega^\perp)$ in the obvious way, we note that $U \in \mathscr{B}_{INV}(\Omega_0 \to \Omega)$ iff it can be written as

$$U = u + v \quad \text{where} \quad u \in \mathscr{B}_{inv}(\Omega_0 \to \Omega) \quad \text{and} \quad v \in \mathscr{B}_{inv}(\Omega_0^\perp \to \Omega^\perp), \tag{1.4}$$

(in this paper, *iff* with double-*f* in stands for *if and only if*). Note also that the set $\mathscr{B}_{INV}(\Omega_0 \to \Omega)$ is the same as $\mathscr{B}_{INV}(\Omega_0^\perp \to \Omega^\perp)$.

If we know an operator $u \in \mathscr{B}_{inv}(\Omega_0 \to \Omega)$ with corresponding inverse $t$, the wanted perturbed eigenvalues of the Hamiltonian $H$ can be found as the eigenvalues of

$$A = tHu = A_0 + \lambda A_1 + \lambda^2 A_2 + \dots \tag{1.5}$$

which is therefore called an *effective* Hamiltonian. In general this is only hermitean, if $u \in \mathscr{B}_{uni}$. When an eigenvalue $E$ of $A$ with eigenvector $|a_0\rangle \in \Omega_0$ has been determined, the real eigenvector corresponding to that $E$ can be determined as

$$|a\rangle = u|a_0\rangle ; \qquad u = u_0 + \lambda u_1 + \lambda^2 u_2 + \dots \tag{1.6}$$

Any transformation $U \in \mathscr{B}_{INV}(\Omega_0 \to \Omega)$ defines a $u = UP_0 \in \mathscr{B}_{inv}(\Omega_0 \to \Omega)$ with corresponding $t = P_0 U^{-1}$ and effective Hamiltonian $A = P_0 U^{-1} HUP_0$.

The pioneering approach is due to Van Vleck [1,2]. The later Foldy-Wouthuysen transformation [3], used it to determine the leading relativistic corrections to Schrödingers non-relativistic Hamiltonian for an electron, can be seen as a special case. This was pointed out by Jørgensen and Pedersen [4] who also simplified general theory [4,5] by use of Foldy and Wouthuysen's notion of *even* and *odd* operators: The arbitrary operator $B$ has a unique decomposition into the sum of an *even* and an *odd* part,

$$B = B_{even} + B_{odd}; \quad \begin{cases} B_{even} = P_0 B P_0 + Q_0 B Q_0 \\ B_{odd} = P_0 B Q_0 + Q_0 B P_0 \end{cases}, \tag{1.7}$$

and $B$ is called *even* if $B_{odd} = 0$ and *odd* if $B_{even} = 0$. One sees that $B$ is *even* iff it commutes with $P_0$ and *odd* if it anticommutes. In the present paper we add a new quite useful observation in terms of the unitary operator

$$R_0 = P_0 - Q_0 \tag{1.8}$$

for the reflection in the subspace $\Omega_0$: $B$ is *even/odd* iff its parity under $R_0$ is positive / negative. See more in appendix A (and see the proof og eqn. (4.1) for a use).

Van Vleck's approach to a unitary $U = e^g \in \mathscr{B}_{UNI}(\Omega_0 \to \Omega)$ can be considered as a systematic method whereby to construct the terms in the anti-hermitean $g = \lambda g_1 + \lambda^2 g_2 + ...$ consecutively so that the terms in the expansion

$$\overline{H} = e^{-g} H e^g = H_0 + \lambda \overline{H}_1 + \lambda^2 \overline{H}_2 + ... \tag{1.9}$$

become *even*. The terms $g_1, g_2, ..$ are not uniquely determined by Van Vlecks procedure, but become so if they are demanded to be *odd*. For reasons we shall return to, the resulting

$$U_c = e^{g_c} = 1 + g_c + \frac{1}{2!} g_c^2 + \frac{1}{3!} g_c^2 + ...; \quad g_c = \lambda g_{c1} + \lambda^2 g_{c2} + .. = (g_c)_{odd}, \tag{1.10}$$

has become known as *the canonical Van Vleck transformation* - hence index "c". Shavitt and Redmond [6] have given a closed formula for the canonical $g_{cn}$ for arbitrary $n$ by combining the notion of *even/odd* with the use of the super-operator formalism.

A pioneering approach due to Bloch [7] differs from Van Vleck's purely algebraic term-by-term procedure by being based on a transformation $\mathscr{U} \in \mathscr{B}_{inv}(\Omega_0 \to \Omega)$ which is defined by an elementary geometric meaning: It is the inverse of the projection map $P_0 : \Omega \to \Omega_0$. More precisely: It is defined by having $t = P_0 P$ as its inverse. This definition leads to a simple recursion relation for the terms in the expansion $\mathscr{U} = P_0 + \lambda \mathscr{U}_1 + \lambda^2 ...$ - and this $\mathscr{U}$ obviously has another attractive characteristic: If

$|a\rangle$ is an eigenvector of the real Hamiltonian, the corresponding eigenvector of Bloch's effective Hamiltonian is the unique vector in $\Omega_0$ which is nearest to $|a\rangle$, namely $|a_0\rangle = P_0|a\rangle$.

Unfortunately Bloch's effective Hamiltonian is not hermitean – only a $u \in \mathscr{B}_{uni}(\Omega_0 \to \Omega)$ ensures that property. A number of subsequent authors have found various seemingly different approaches whereby to correct for this. Des Cloizeux [8] was probably the first to do so and it was for his $u_c \in \mathscr{B}_{uni}(\Omega_0 \to \Omega)$ that Klein [9] introduced the term "canonical". He did so after having found that it is uniquely characterized by the following intuitively very satisfactory property: The orthonormal eigenvectors of the effective Hamiltonian $A_c = u_c^\dagger H u_c$ deviate (in a precise least square sense) as little as possible from those of the real Hamiltonian. Subsequently Klein used this characteristic to show that the canonical $u_c$ is the same as $U_c P_0$ with Van Vleck's $U_c$ from (1.10). His proof, however, is not quite simple and later Jørgensen [10] found that it fails on some treacherous details. By then Jørgensen [11] had found another theorem of uniqueness (derived more simply in sec. 5 of the present paper): There exists one, and only one $u \in \mathscr{B}_{uni}(\Omega_0 \to \Omega)$ for which $P_0 u$ is positive (*i.e.* hermitean with no negative eigenvalues). This theorem reveals almost immediately that $u_c$ is identical to $U_c P_0$. See Jørgensen [11] for other examples on $u$'s which in the same way turn out to be identical to $u_c$. Brandow [12] gives still further examples as well as an elaborate correct version of Klein's defective use of his minimum-principle.

With certain simple definitions and lemmas collected in appendix A, the present paper proceeds as follows. Sec. 2 presents a new formulation of the basic theory in which the simple operator $S = PP_0 + QQ_0$ from $\mathscr{B}_{INV}(\Omega_0 \to \Omega)$ is the central building stone. It is shown to have a property which makes certain algebraic manipulations much easier : It is *normal* (commutes with its own adjoint ). Hence an operator $\theta$ for an angle can be defined by

$$\cos\theta = |S| \; ; \; |S|^2 \equiv \begin{cases} S^\dagger S = P_0 P P_0 + Q_0 Q Q_0 \\ SS^\dagger = PP_0 P + QQ_0 Q \end{cases}, \tag{1.11}$$

and the demand that all its eigenvalues are in the interval $[0, \frac{\pi}{2}[$. We then show that the canonical $U_c$ can be written as

$$U_c = e^{g_c} = \cos\theta + c_4 \sin\theta; \quad g_c = \theta c_4 = c_4 \theta, \tag{1.12}$$

where $c_4$ is an operator for "a rotation of angle $\frac{\pi}{2}$ around the null space" for $\theta$ - that is the space of intersections $\omega^\perp \equiv (\Omega_0 \cap \Omega) \cup (\Omega_0^\perp \cap \Omega^\perp)$ (= the eigenspace of $\cos^2\theta = |S|^2$ corresponding to the eigenvalue 1). So $U_c$ rotates vectors in the "plane" $\omega = (\omega^\perp)^\perp$ while vectors in the "normal" $\omega^\perp$ are left unchanged.

In sec. 3 we show how this somewhat intuitive description becomes concrete in the simple case where $\Omega_0$ and $\Omega$ are two-dimensional planes in the 3-dimensional space of elementary geometry. Fig. 3.1 is central by showing how $U_c$ can be considered as a rotation of an acute angle $\theta_1$ around a z-axis chosen along the line of intersection $\Omega_0 \cap \Omega$. As well known from elementary geometry, this rotation is the same as the product $R_m R_0$ where $R_m$ is the reflection in the mirror plane $\Omega_m$ which bisects the angle $\theta_1$. The intention of the quite detailed sec. 3 is to give a preliminary clear understanding of the geometrical ideas which are generalized in sec. 4. It turns out that most of the formulas for the elementary 3-dimensional case can be used as they stand also in general case. Of course a little extra care is needed to explain the meanings of the operators and subspaces. Thus $\omega$ in the 3-dimensional case is the 2-dimensional eigenspace of $\theta$ corresponding to the eigenvalue $\theta_1$. In the general case, $\theta$ has several such 2-dimensional eigenspaces. The central Fig.4.1, which illustrates the action of $U_c$ within one of these, is easily understood if one understands the action of $U_c$ within the plane $\omega$ in fig. 3.1.

In sec. 3 Klein's theorem is found by a very simple observation. In sec. 5 this is generalized to give the theorem in various refined forms. Its characteristic minimum-property seems to be rather unrelated to Jørgensen's $P_0 u$ being positive. But in sec. 6 we start by using the latter for a new very simple direct derivation of Klein's result. On the background of this we then look into Klein's original proof and we substantially simplify Brandow's elaborate variational proof of $U_c$'s minimal-effect. We also sim-plify and extend an earlier proof due to Jørgensen [13] in which the transformation operators $u$ are considered as vectors with a length (norm). Klein's theorem then asserts that no other $u \in \mathscr{B}_{uni}(\Omega_0 \to \Omega)$ comes as close to $P_0$ as $u = u_c$. An extension asserts that that no other $u$ comes as close to Bloch's $\mathscr{U}$ as the canonical $u = u_c$. The main intention of the section is to give a background on which to appreciate what has been achieved in the previous sections. In the final sec. 7 we sum up and conclude.

## 2: Basics of the canonical operators

Perturbation theory is designed for cases where the spaces $\Omega_0$ and $\Omega$ are "fairly close" to each other. Following Bloch [7] we express this by assuming that $\Omega_0$ and $\Omega$ are *non-orthogonal*, hereby meaning that no vector in one of the spaces is orthogonal to all vectors in the other (see also app. B). Or, equivalently: If $|x_0\rangle$ is a non-zero vector in $\Omega_0$ then $|x\rangle = P|x_0\rangle$ is a non-zero in vector $\Omega$. And if $|x\rangle$ is a non-zero vector in $\Omega$ then $|x_0\rangle = P_0|x\rangle$ is a non-zero vector $\Omega_0$. In such a case $\langle x_0|P_0PP_0|x_0\rangle = \||P|x_0\rangle\|^2$ is positive for all $|x_0\rangle \in \Omega_0$, implying that $P_0PP_0$ has a likewise positive inverse $(P_0PP_0)^{-1}$ on $\Omega_0$ (satisfying $(P_0PP_0)^{-1}(P_0PP_0) = (P_0PP_0)(P_0PP_0)^{-1} = (P_0PP_0)^0 = P_0$). Hence we can define a positive operator $(P_0PP_0)^r$ on $\Omega_0$ for *any* real number $r$ - and define $(PP_0P)^r$ similarly on $\Omega$.

The non-orthogonal property of $\Omega_0$ and $\Omega$ can be expressed equivalently by saying that the two subspaces $\Omega_0 \cap \Omega^\perp$ and $\Omega_0^\perp \cap \Omega$ are empty (except for the null-vector). This property is obviously symmetric under the interchange $(\Omega_0, \Omega) \leftrightarrow (\Omega_0^\perp, \Omega^\perp)$. Hence $\Omega_0$ and $\Omega$ are non-orthogonal iff the orthogonal complements $\Omega_0^\perp$ and $\Omega^\perp$ are so. It follows that $\Omega_0$ and $\Omega$ are non-orthogonal iff the operator

$$S = PP_0 + QQ_0 \tag{2.1}$$

is invertible. From now on that is assumed to be the case unless something else is asserted explicitly. It then follows directly from the obvious relation $SP_0 = SP$ that $S \in \mathscr{B}_{INV}(\Omega_0 \to \Omega)$. This operator $S$ is fundamental for the theory of the present paper. In appendix D we show that it is *normal* (commutes with its adjoint $S^\dagger$) and thus has the properties described in appendix C. In accordance with (D6) and (D7) we introduce the "angle-operator" $\theta$ defined by

$$\cos^2\theta = |S|^2 = \begin{cases} S^\dagger S = P_0PP_0 + Q_0QQ_0 \\ SS^\dagger = PP_0P + QQ_0Q \end{cases} \tag{2.2}$$

or equivalently by

$$\sin^2\theta = 1 - |S|^2 = \begin{cases} P_0QP_0 + Q_0PQ_0 \\ PQ_0P + QP_0Q \end{cases} \text{ or by} \tag{2.3}$$

$$\cos\theta \sin\theta = |S_{odd}| \tag{2.4}$$

together with the demand that arbitrary eigenvalue $\theta_\mu$ should be an angle in the interval $0 \leq \theta_\mu < \frac{\pi}{2}$, now with $\frac{\pi}{2}$ excluded because the invertible $|S| = \cos\theta$ cannot have zero as an eigenvalue. Hence the relation (D5) can here be written as

$$\sin\theta = \frac{|S_{odd}|}{|S|}. \tag{2.5}$$

By the definition in (2.2) we have

$$\cos^r \theta = (P_0 P P_0)^{\frac{r}{2}} + (Q_0 Q Q_0)^{\frac{r}{2}} = (P P_0 P)^{\frac{r}{2}} + (Q Q_0 Q)^{\frac{r}{2}} \tag{2.6}$$

for all positive real numbers $r$. With the precaution from the beginning of this section, the result is also valid for negative real $r$.

The polar part (see (C7)) of $S$, i.e. the unitary operator

$$U_c = \frac{S}{|S|} = \begin{cases} S|S|^{-1} = P(P_0 P P_0)^{-\frac{1}{2}} + Q(Q_0 Q Q_0)^{-\frac{1}{2}} \\ |S|^{-1} S = (P P_0 P)^{-\frac{1}{2}} P_0 + (Q Q_0 Q)^{-\frac{1}{2}} Q_0 \end{cases} \tag{2.7}$$

belongs to $\mathscr{B}_{UNI}(\Omega_0 \to \Omega)$ ( because $U_c P_0 = P U_c$ ). Hence

$$u_c = \frac{S}{|S|} P_0 = \begin{cases} P(P_0 P P_0)^{-\frac{1}{2}} \\ (P P_0 P)^{-\frac{1}{2}} P_0 \end{cases} \tag{2.8}$$

belongs to $\mathscr{B}_{uni}(\Omega_0 \to \Omega)$. Note that one obtains $S^\dagger$, $U_c^\dagger$ and $u_c^\dagger$ from $S$, $U_c$ and $u_c$ by interchanging the spaces $\Omega_0$ and $\Omega$ ( that is, by interchanging $(P_0, Q_0)$ and $(P, Q)$ ). We write this result as

$$\Omega_0 \leftrightarrow \Omega \Rightarrow (S, U_c, u_c) \leftrightarrow (S_c^\dagger, U_c^\dagger, u_c^\dagger) \tag{2.9}$$

Jørgensen's theorem of uniqueness [11]( derived more simply in the start of sec. 5 of the present paper) asserts that: If $U \in \mathscr{B}_{UNI}(\Omega_0 \to \Omega)$ and $u \in \mathscr{B}_{uni}(\Omega_0 \to \Omega)$ then

$$U_{even} \text{ is positive} \Leftrightarrow U_{even} = |S| \Leftrightarrow U = U_c \tag{2.10a}$$

$$P_0 u \text{ is positive} \Leftrightarrow P_0 u = |S| P_0 \Leftrightarrow u = u_c. \tag{2.10b}$$

Aiming at the intuitive idea that $U_c$ can be considered as a rotation around the "normal" $\omega^\perp$ to a certain "plane" $\omega$, we define the *space of intersections*

$$\omega^\perp = (\Omega_0 \cap \Omega) \oplus (\Omega_0^\perp \cap \Omega^\perp), \tag{2.11}$$

with the orthogonal complement $\omega = (\omega^\perp)^\perp$. Let $P_\omega$ and $Q_\omega = 1 - P_\omega$ be the projectors on $\omega$ and $\omega^\perp$ respectively. From (2.2) one sees that $\omega^\perp$ is the eigenspace of $|S| = \cos\theta$ corresponding to the eigenvalue 1. By (2.3) that is the same as is the null-space for $\sin\theta$ - and hence for $\theta$ and for $|S_{odd}| = \cos\theta \sin\theta$. We therefore have

$$\theta = P_\omega \theta, \quad \sin\theta = P_\omega \sin\theta \quad \text{and} \quad \cos\theta = Q_\omega + P_\omega \cos\theta \tag{2.12}$$

$(= Q_\omega \cos(0) + P_\omega \cos\theta)$.

Since $S_{odd}$ equals $S_A$, it is antihermitean - and thereby normal. And since $\omega$ is orthogonal to its null-space then $S_{odd}$ and $|S_{odd}|$, considered as operators on $\omega$, have inverses on that space and we can give $S_{odd}$ by its polar decomposition – that is we can write it as

$$S_{odd} = |S_{odd}| c_4 = c_4 |S_{odd}| \quad \text{where} \quad c_4 = P_\omega \frac{S_{odd}}{|S_{odd}|} P_\omega \tag{2.13}$$

is antihermitean and unitary on $\omega$ (*i.e.* $c_4 c_4^\dagger = c_4^\dagger c_4 = P_\omega$). Actually it satisfies the demands from (A2) for a quarter turn within $\omega$:

$$c_4^2 = -c_4 c_4^\dagger = -P_\omega, \ c_4^3 = -P_\omega c_4, \ c_4^4 = P_\omega, \ c_4^5 = c_4, \ \ldots. \tag{2.14}$$

Note that with the (important) exception that $P_0$ does not commute with $P$ then all the operators $P_0$, $P, P_\omega, c_4, \theta, \sin\theta \ |S| = \cos\theta, \ |S_{even}| = \cos\theta$ and $|S_{odd}| = \sin\theta \cos\theta$ commute among each other.

We now prove that the $U_c$ from (2.7) can be written as

$$U_c = \cos\theta + c_4 \sin\theta = D(\theta) \tag{2.15}$$

where $D(\theta)$ is obtained by letting the operator $\theta$, which commutes with $c_4$, take the place of the angle $\varphi$ in the rotation operator $D(\varphi)$ from (A5). That is,

$$D(\theta) = e^{c_4\theta} = \begin{cases} Q_\omega + P_\omega(\cos\theta + c_4\sin\theta) \\ Q_\omega + P_\omega e^{c_4\theta} \end{cases}. \tag{2.16}$$

where $P_\omega, c_4$ and $\theta$ commute among each other.

*Proof*: The first expression in (2.15) is obtained by use of (2.5) in the rearrangement

$$U_c = \frac{S_{even}}{|S|} + \frac{S_{odd}}{|S|} = \cos\theta + P_\omega \frac{S_{odd}}{|S_{odd}|} \frac{|S_{odd}|}{|S|}. \tag{2.17}$$

with $c_4$ from (2.13). To verify the next, just note that since $\theta$ commutes with $c_4$, it can take the place of $\varphi$ in the derivation of the formula (A8). This gives the expressions in (2.16). Use finally eqn. (2.12) for $\cos\theta$ Q.E.D.

It follows from (2.15) that

$$U_{c,even} = U_{c,H} = \cos\theta \quad \text{and} \quad U_{c,odd} = U_{c,A} = c_4\sin\theta \tag{2.18}$$

# 3: Geometry I: The case of three dimensions.

Wishing to discuss geometric aspects of the canonical Van Vleck transformation, we start by considering the elementary example where $\Omega_0$ and $\Omega$ are 2-dimensional subspaces in a 3-dimensional Hilbert space. See fig. 3.1.

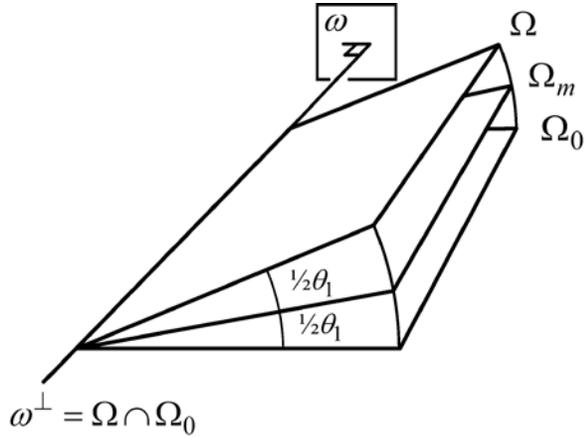
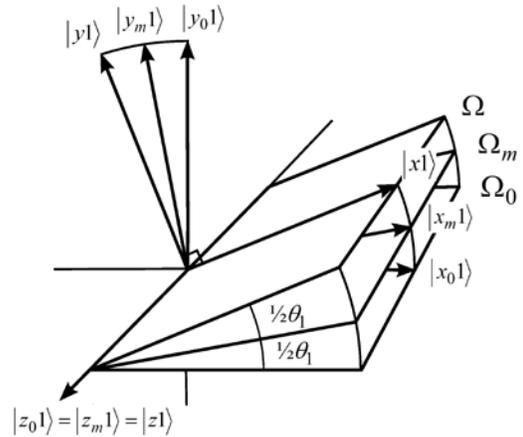

**Figure 3.1a:** Illustrates the case where $\Omega_0$ and $\Omega$ are 2-dimensional subspaces in 3-dimensional Hilbert space with an acute angle $\theta_1 > 0$ between them. The reason for the index "1" (and for the number "1" in the next figure's basis vectors $|..1\rangle$) becomes clear as the theory proceeds. Index "$m$" on $\Omega_m$, the plane which bisects the angle $\theta_1$, stands for "mirror", indicating that the reflection $R_m$ in this plane maps $\Omega_0$ on $\Omega$ and vice versa.
The subspace orthogonal to the line of intersection $\Omega_0 \cap \Omega$ is called $\omega$.

**Figure 3.1b:** Three frames of orthonormal axes are now shown together with the planes from fig. 3.1a: $K_0$ with axes $|\alpha_0 1\rangle$; $\alpha = x, y, z$, $K_m$ with axes $|\alpha_m 1\rangle$ and $K$ with axes $|\alpha 1\rangle$.
Each of the three sets $(|x_0 1\rangle, |y_0 1\rangle)$, $(|x 1\rangle, |y 1\rangle)$ and $(|x_m 1\rangle, |y_m 1\rangle)$ form an orthonormal basis in the plane $\omega$.
The line of intersection $\omega^\perp = \Omega_0 \cap \Omega$ is spanned by $|z_0 1\rangle = |z 1\rangle = |z_m 1\rangle$.

Intuitively fig. 3.1a immediately suggests that if some mapping which transforms $\Omega_0$ to $\Omega$ deserves to be called canonical, it should be the rotation $D(\varphi)$ of angle $\varphi = \theta_1$ around the line of intersection. Also the reflection $R_m$ springs to mind. So, by a well known result for two consecutive reflections, the immediate expectation is that

$$U_c = D(\theta_1) = R_m R_0 \quad \text{and} \quad u_c = U_c P_0 = R_m P_0 \tag{3.1}$$

where $R_0 = P_0 - Q_0$ is the reflection in $\Omega_0$. To consider these expectations further, we introduce some terminology.

It is obvious how the three sets of orthonormal vectors in fig. 3.1b are such constructed that the projectors on the four subspaces $\omega, \Omega_0, \Omega_m, \Omega$ and their orthogonal complements can be written as

$$P_\omega = \begin{cases} |x_01\rangle\langle x_01| + |y_01\rangle\langle y_01| \\ |x_m1\rangle\langle x_m1| + |y_m1\rangle\langle y_m1| \\ |x1\rangle\langle x1| + |y1\rangle\langle y1| \end{cases} \text{ and } Q_\omega = \begin{cases} |z_01\rangle\langle z_01| \\ |z_m1\rangle\langle z_m1|, \\ |z1\rangle\langle z1| \end{cases}$$

$$P_0 = |z_01\rangle\langle z_01| + |x_01\rangle\langle x_01|, \quad Q_0 = |y_01\rangle\langle y_01|,$$

$$P_m = |z_m1\rangle\langle z_m1| + |x_m1\rangle\langle x_m1|, \quad Q_m = |y_m1\rangle\langle y_m1|,$$

$$P = |z1\rangle\langle z1| + |x1\rangle\langle x1|, \quad Q = |y1\rangle\langle y1|, \tag{3.2}$$

where $|z_01\rangle = |z_m1\rangle = |z1\rangle$. The reflections $R_\omega, R_0, R$ and $R_m$ in $\omega, \Omega_0, \Omega$ and $\Omega_m$ are given by $R_\omega = P_\omega - Q_\omega$, etc. One sees directly from the figure that $R_m$ interchanges $(P_0, Q_0)$ and $(P, Q)$:

$$R_m(P_0, Q_0)R_m = (P, Q) \text{ and } R_m(P, Q)R_m = (P_0, Q_0). \tag{3.3}$$

Observations like $P_0 D(\theta_1)|x_01\rangle = P_0|x1\rangle = \cos\theta_1|x_01\rangle$ show that $D_{even} = P_0 D(\theta_1)P_0 + Q_0 D(\theta_1)Q_0$ has the three orthonormal eigenvectors $|x_01\rangle, |y_01\rangle$ and $|z_01\rangle$ with corresponding eigenvalues $\cos\theta_1$, $\cos\theta_1$ and 1. Since these are all positive, $D_{even}$ is hermitean as well as positive and it follows from the theorem of uniqueness in (2.10a) that $D(\theta_1)$ in (3.1) is the canonical $U_c$ - as expected.

Since $P_\omega$ and $Q_\omega$ are *even* (invariant under $R_0$), so is the angle operator defined by

$$\theta = \theta_1 P_\omega \quad (= \theta_1 P_\omega + 0 \cdot Q_\omega), \tag{3.4}$$

as well as the operators

$$\cos\theta = \cos\theta_1 P_\omega + Q_\omega \quad (= \cos\theta_1 P_\omega + \cos(0)Q_\omega),$$

$$\sin\theta = \sin\theta_1 P_\omega. \tag{3.5}$$

We can now verify the result postulated in (1.11):

$$\cos^2\theta = P_0 P P_0 + Q_0 Q Q_0 = P P_0 P + Q Q_0 Q \tag{3.6}$$

*Proof*: Use that two hermitean operators are identical iff they have the same eigenvectors and eigenvalues. The identities aimed at then follows from results like $P_0 P P_0 |x_0 1\rangle = \cos\theta_1 P_0 |x1\rangle = \cos^2\theta_1 |x_0 1\rangle$, $P_0 P P_0 |z_0 1\rangle = |z_0 1\rangle$ and $Q Q_0 Q |y1\rangle = \cos\theta_1 Q |y_0 1\rangle = \cos^2\theta_1 |y1\rangle$, seen directly from the geometry in fig. 3.1. Q.E.D. The identities

$$\sin^2\theta = P_0 Q P_0 + Q_0 P Q_0 = P Q_0 P + Q P_0 Q \tag{3.7}$$

can be verified analogously – or by use $\sin^2\theta = 1 - \cos^2\theta$ and results like $P_0 P P_0 = P_0(1-Q)P_0 = P_0 - P_0 Q P_0$.

The rotation $D(\varphi)$ can be given by the formula in (A5),

$$D(\varphi) = e^{c_4 \varphi} = \begin{cases} Q_\omega + P_\omega(\cos\varphi + c_4 \sin\varphi) \\ Q_\omega + P_\omega e^{c_4 \varphi} \end{cases} \tag{3.8}$$

where $c_4$ is the operator which replaces the arbitrary vector $|a\rangle$ with the vector product $|z1\rangle \times |a\rangle$.

*Proof*: Since $c_4$ has the properties of a *quarter turn within* $\omega$ described in (A2), the formula (A5) applies. To verify the action of $D(\varphi)$ explicitly, just note that the upper expression in (3.8) leaves $|z_0 1\rangle$ unchanged, transforms $(|x_0 1\rangle$ to $\cos\varphi |x_0 1\rangle + \sin\varphi |y_0 1\rangle$ and $|y_0 1\rangle$ to $-\sin\varphi |x_0 1\rangle + \cos\varphi |y_0 1\rangle$. Q.E.D. Since the vector product is only defined in three dimensions, it is not useful for generalization. We therefore observe that $c_4$ also can given be given by the expression

$$c_4 = |y_0 1\rangle\langle x_0 1| - |x_0 1\rangle\langle y_0 1| \quad = \frac{P P_0 - P_0 P}{\cos\theta_1 \sin\theta_1} \tag{3.9}$$

which is obviously antihermitean, *odd* and of the form $c_4 = P_\omega c_4 P_\omega$. *Proof*: The first expression is obvious. To check the second, write $P P_0 - P_0 P$ as $|x1\rangle\langle x1|x_0 1\rangle\langle x_0 1|$ – adjoint where $\langle x1|x_0 1\rangle = \cos\theta_1$ and $|x1\rangle = \cos\theta_1 |x_0 1\rangle + \sin\theta_1 |y_0 1\rangle$. Q.E.D.

Since $\theta$ commutes with $c_4$ it can take the place of the number $\varphi$ in the derivation of the result (3.8). Doing so, and using (3.5), we see that $U_c = D(\theta_1)$ can be written as

$$U_c = D(\theta) = e^{c_4 \theta} = \cos\theta + c_4 \sin\theta \tag{3.10}$$

Furthermore: Tacitly agreeing always to read $c_4$ as $P_\omega c_4 P_\omega$, the angle $\theta_1$ in the denominator in (3.9) can be replaced with the operator $\theta$. We then get

$$c_4 = \frac{PP_0 - P_0 P}{\cos\theta \sin\theta} = \frac{QQ_0 - Q_0 Q}{\cos\theta \sin\theta}. \qquad (3.11)$$

The second expression follows from the first when inserting $P = 1 - Q$ and $P_0 = 1 - Q_0$.

## *Klein's least square characterization*

For given $u \in \mathcal{B}_{uni}(\Omega_0 \to \Omega)$, let $|\eta_0\rangle$; $\eta = 1,2$, be an orthonormal basis in the plane $\Omega_0$. Then the positive number $\varepsilon$ defined by

$$\varepsilon^2 = \Sigma_\eta \big\| |\eta\rangle - |\eta_0\rangle \big\|^2 \; ; \; |\eta\rangle = u|\eta_0\rangle, \qquad (3.12)$$

is a measure of how much $u$ changes these basis vectors when transforming them to a basis in $\Omega$. Writing $\big\|(u - P_0)|\eta_0\rangle\big\|^2$ as $\langle \eta_0 |(u^\dagger - P_0)(u - P_0)|\eta_0\rangle$ we get

$$\varepsilon^2 = \mathrm{Tr}\{2P_0 - P_0(u + u^\dagger)P_0\} = 2\mathrm{Tr}\{P_0 - P_0 u_H P_0\}. \qquad (3.13)$$

where the trace is independent of the basis in which it is evaluated. Hence $\varepsilon$ is a function $\varepsilon(u)$ of $u$ only. Choosing $\{|z_0 1\rangle, |x_0 1\rangle\}$ as the basis we get

$$\varepsilon(u)^2 = \big\| u|z_0 1\rangle - |z_0 1\rangle \big\|^2 + \big\| u|x_0 1\rangle - |x_0 1\rangle \big\|^2 \qquad (3.14)$$

One sees directly from the figure that no unit vector in $\Omega$ is as close to / distant from $|z_0 1\rangle$ as $|z 1\rangle / -|z 1\rangle$. And no unit vector in $\Omega$ is as close to / distant from $|x_0 1\rangle$ as $|x 1\rangle / -|x 1\rangle$. In terms of the lengths

$$d_{1\mp} = \big\| |x1\rangle \mp |x_0 1\rangle \big\| = \begin{cases} 2\sin\frac{\theta_1}{2} \\ 2\cos\frac{\theta_1}{2} \end{cases} \qquad (3.15)$$

(which satisfy $d_{1-}^2 + d_{1+}^2 = 4$) and with $\varepsilon(u_c)$ abbreviated to just $\varepsilon_c$ we therefore have

$$\varepsilon_c^2 = d_{1-}^2 \leq \varepsilon(u)^2 \leq 4 + d_{1+}^2 = 8 - \varepsilon_c^2 \qquad (3.16)$$

where minimum/maximum is attained iff $u = u_c / u = -u_c$. This result is a more elaborate version of Klein's least squares characterization of the canonical $u_c$. See more in sec. 5. It can be elaborated still further:

It is obvious how the definition in (3.12) should be modified to give a measure $E(U)$ of how much a given $U \in \mathscr{B}_{UNI}(\Omega_0 \to \Omega)$ changes the vectors in a orthonormal basis $|\eta_0\rangle$; $\eta = 1,2,3$, of the full 3-dimensional space. The rearrangement

$$E(U)^2 = \Sigma_\eta \|U|\eta_0\rangle - |\eta_0\rangle\|^2 = 2\text{Tr}\{1 - U_H\} \tag{3.17}$$

shows that $E(U)$ is independent of how the basis vectors $|\eta_0\rangle$ are chosen. Using $|\alpha_0 1\rangle$; $\alpha = x, y, z$, we get

$$E(U)^2 = \|U|z_0 1\rangle - |z_0 1\rangle\|^2 + \|U|x_0 1\rangle - |x_0 1\rangle\|^2 + \|U|y_0 1\rangle - |y_0 1\rangle\|^2 \tag{3.18}$$

where $U|\alpha_0 1\rangle$ belongs to $\Omega$ when $\alpha = z, x$ and belongs to $\Omega^\perp$ when $\alpha = y$. With $\varepsilon_c^2 \equiv d_{1-}^2$ as in (3.16) one finds

$$2\varepsilon_c^2 \leq E(U)^2 \leq 4 + 2d_{1+}^2 = 12 - 2\varepsilon_c^2 \tag{3.19}$$

where the minimum/maximum is attained if, and only if, $U = U_c / U = -U_c$.

*Proof*: The two first terms in (3.18) can be evaluated exactly as in the same terms in (3.14). And as to the extra last term, the result (3.15) remains valid as it stands when $|x1\rangle \mp |x_0 1\rangle$ is replaced with $|y1\rangle \mp |y_0 1\rangle$. *Q.E.D.*

*A warning about the uniqueness of $g = g_c$ in $e^g = e^{g_c}$*

As mentioned in connection with of eqn. (1.10), the terms in Van Vleck's expansion $g = \lambda g_1 + \lambda^2 g_2 + ...$ are uniquely determined if they are demanded to be *odd*. A warning goes with this assertion: The equation $U_c = e^g$ have solutions $g$ which are not *odd* and it has *odd* solutions which differ from $g_c = c_4 \theta$.

This warning is substantiated at the end of app. C with the present 3-dimensinal case as an example. It becomes clear, that the ambiguities disappear with the extra demand that $g \to 0$ when $\Omega \to \Omega_0$ - as it is the case when $\lambda \to 0$ in the perturbation problem (1.1).

## 4: Geometry II: The general case

In the 3-dimensional case illustrated in fig. 3.1, the mirror space $\Omega_m$ is obtained by subjecting $\Omega_0$ to the rotation of angle $\frac{\theta_1}{2}$ around the line of intersection. For the generalization of the results (3.1) we now show that

$$U_c = R_m R_0 \text{ where } R_m \equiv e^{\frac{g_c}{2}} R_0 e^{-\frac{g_c}{2}} \tag{4.1}$$

is the reflection in the "mirror plane" $\Omega_m$, obtained from $\Omega_0$ by the unitary $e^{\frac{g_c}{2}}\ (= D(\frac{\theta}{2}))$.

*Proof*: Just insert $R_m$ in $R_m R_0$ and observe that $R_0 e^{-\frac{g_c}{2}} R_0 = e^{-R_0 \frac{g_c}{2} R_0}$ equals $e^{\frac{g_c}{2}}$. *Q.E.D.*

Writing that canonical $U_c$ as $D(\theta)$ in (2.16) one sees that $U_c$ has no effect on vectors in the space $\omega^\perp$ of intersections (on which $Q_\omega$ is the unity operator). To describe the action of $U_c$ it is therefore sufficient to study what it does to vectors in $\omega$. To do so, we first write this space as

$$\omega = \omega_{x0} \oplus \omega_{y0} \text{ where } \omega_{x0} = \Omega_0 \cap \omega \text{ and } \omega_{y0} = \Omega_0^\perp \cap \omega. \tag{4.2}$$

From an orthonormal basis in $\Omega_0$ of eigenvectors for $P_0 P P_0$, we take those away which belongs to the intersection $\Omega_0 \cap \Omega$; *i.e.* those corresponding to the eigenvalue 1. The remaining then form an orthonormal basis $|x_0 \mu\rangle$; $\mu = 1,2,...$ in $\omega_{x0}$ such that

$$\cos\theta |x_0\mu\rangle = (P_0 P P_0)^{\frac{1}{2}} |x_0\mu\rangle = \cos\theta_\mu |x_0\mu\rangle \tag{4.3}$$

where $0 < \theta_\mu < \frac{\pi}{2}$ for all $\mu$. If $\Omega_0$ is of finite dimension $g$, the number of $\mu'$s equals $n_0 = g - g_0$ where $g_0$ is the dimension of $\Omega_0 \cap \Omega$. Next we introduce the vectors $|y_0\mu\rangle \equiv c_4 |x_0\mu\rangle$ and show that they form an orthonormal basis in $\omega_{y0}$ such that

$$\cos\theta |y_0\mu\rangle = (Q_0 Q Q_0)^{\frac{1}{2}} |y_0\mu\rangle = \cos\theta_\mu |y_0\mu\rangle. \tag{4.4}$$

*Proof*: The identities in (4.4) are obtained by acting on those in (4.3) with $c_4$ which commutes with $\cos\theta$ and is *odd*. The vectors $|y_0\mu\rangle$ are orthonormal because $c_4$ is unitary on $\omega$. To see that they span the whole of $\omega_{y0}$, *i.e.* that the arbitrary $|b\rangle \in \omega_{y0}$ can be written as $\Sigma_\mu b_\mu |y_0\mu\rangle$, write first $|a\rangle \equiv c_4^\dagger |b\rangle$ as $\Sigma_\mu a_\mu |x_0\mu\rangle$. We then get $|b\rangle = c_4 |a\rangle = \Sigma_\mu a_\mu |y_0\mu\rangle$. Q.E.D.

It follows that $\omega$ can be written as the sum

$$\omega = \Sigma_\mu \oplus \omega_\mu \tag{4.5}$$

where $\omega_\mu$ is spanned by the two orthonormal eigenvectors $|x_0\mu\rangle$ and $|y_0\mu\rangle = c_4|x_0\mu\rangle$ of $\theta$, both corresponding to the same eigenvalue $\theta_\mu$,

$$\theta|x_0\mu\rangle = \theta_\mu |x_0\mu\rangle \quad \text{and} \quad \theta|y_0\mu\rangle = \theta_\mu |y_0\mu\rangle. \tag{4.6}$$

We shall think of $\omega_\mu$ as a plane where $|x_0\mu\rangle$ and $|y_0\mu\rangle$ are the basis vectors in a usual two-dimensional coordinate system as illustrated in fig. 4.1. One sees that $\omega_\mu$ *reduces* ( see note at the start of app. A) $c_4$ and $\theta$ - and hence also reduces operators such as $D(\varphi)$ and $D(\theta)$ from (2.15) and (2.16). The action of $D(\varphi)$ on a vector in $\omega_\mu$, considered as a plane, is to subject it to a rotation of angle $\varphi$, the positive direction being from $|x_0\mu\rangle$ towards $|y_0\mu\rangle$. *Proof*: $D(\varphi)|\alpha_0\mu\rangle$ equals $\cos\varphi |x_0\mu\rangle + \sin\varphi |y_0\mu\rangle$ when $\alpha = x$ and equals $-\sin\varphi |x_0\mu\rangle + \cos\varphi |y_0\mu\rangle$ when $\alpha = y$. Q.E.D. For $U_c = D(\theta)$ one finds

$$U_c = D(\theta)|a\rangle = D(\theta_\mu)|a\rangle \text{ for any } |a\rangle \in \omega_\mu \tag{4.7}$$

Since all vectors in $\omega_\mu$ are eigenvectors of $\theta$ corresponding to the eigenvalue $\theta_\mu$ so are the orthonormal vectors

$$|x\mu\rangle = U_c|x_0\mu\rangle = \cos\theta_\mu |x_0\mu\rangle + \sin\theta_\mu |y_0\mu\rangle \in \Omega \text{ and}$$

$$|y\mu\rangle = U_c|y_0\mu\rangle = -\sin\theta_\mu |x_0\mu\rangle + \cos\theta_\mu |y_0\mu\rangle \in \Omega^\perp, \tag{4.8}$$

also illustrated in fig. (4.1). In analogy with (4.3) and (4.4), these vectors satisfy

$$\cos\theta |x\mu\rangle = (PP_0P)^{\frac{1}{2}}|x\mu\rangle = \cos\theta_\mu |x\mu\rangle \tag{4.9}$$

$$\cos\theta|y\mu\rangle = (QQ_0Q)^{\frac{1}{2}}|y\mu\rangle = \cos\theta_\mu|y\mu\rangle \qquad . \tag{4.10}$$

To prevent overloading fig. 4.1 we omitted the vectors $|\alpha_m\mu\rangle = D(\frac{\theta}{2})|\alpha_0\mu\rangle$; $\alpha = x, y,$ which generalize the vectors $|\alpha_m 1\rangle$ in fig. 3.1.

We leave it for the reader to generalize the subsection *Note on the "natural" exponential form of $U_c$* at the end of section 3 about the 3-dimensional case. It is straightforward.

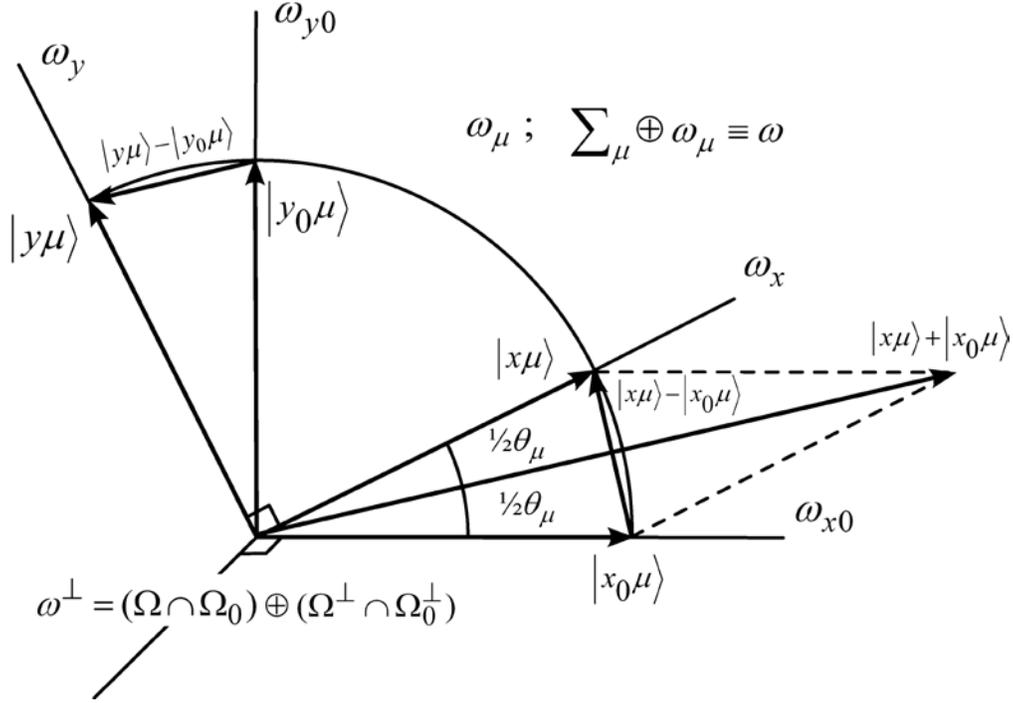

**Figure 4.1:** In the 3-dimensional case illustrated in fig. 3.1, the canonical Van Vleck transformation $U_c$ is a rotation of a certain angle acute $\theta_1$ around the line of intersection $\omega^\perp$ for the two planes, normal to the plane $\omega$. In the general case, the 1-dimensional line of intersection in fig. 3.1 is replaced with the *space of intersections*

$$\omega^\perp = (\Omega_0 \cap \Omega) \oplus (\Omega_0^\perp \cap \Omega^\perp)$$

and the plane $\omega$ in fig. 3.1 is replaced with the subspace which has this $\omega^\perp$ as its orthogonal complement. In fig. 3.1, $\omega$ is a single 2-dimensional subspace with the orthonormal axes $|\alpha_0 1\rangle$; $\alpha = x, y$, as well as with the axes $|\alpha_m 1\rangle$ and $|\alpha 1\rangle$ obtained by rotations of acute angles $\tfrac{1}{2}\theta_1$ and $\theta_1$, respectively. In the general case $\omega$ is a sum

$$\omega = \Sigma_\mu \oplus \omega_\mu \ ; \ \mu = 1, 2, \ldots,$$

of such 2-dimensional subspaces where $\omega_\mu$ is spanned by the two orthonormal axes $|\alpha_0 \mu\rangle$; $\alpha = x, y$, as well as by the axes $|\alpha_m \mu\rangle$ and $|\alpha \mu\rangle$ obtained from these by rotations of acute angles $\tfrac{1}{2}\theta_\mu$ and $\theta_\mu$, respectively.

In the figure, the vectors $|\alpha_0 1\rangle$ and $|\alpha 1\rangle$ are shown as arrows while the subspaces $\omega_{\alpha 0}$ and $\omega_\alpha$ to which they belong are represented by a lines. The unit vectors $|x_m \mu\rangle$ and $|y_m \mu\rangle$ are given by

$$\left.\begin{matrix}|x_m \mu\rangle\\|y_m \mu\rangle\end{matrix}\right\} = \frac{|x\mu\rangle \pm |x_0 \mu\rangle}{d_{\mu\pm}} \ ; \ d_{\mu\pm} = \||x\mu\rangle \pm |x_0\mu\rangle\| = \||y\mu\rangle \pm |y_0\mu\rangle\| = \begin{cases} 2\cos\dfrac{\theta_\mu}{2} \\ 2\sin\dfrac{\theta_\mu}{2} \end{cases},$$

To prevent overloading the figure, these unit vectors are not shown explicitly. The lengths $d_{\mu\pm}$ turn out to have a clear meaning when Klein's least square characterization is formulated as in sec. 5.

## 5: The two theorems of uniqueness. Extension of Klein's.

The arbitrary $u \in \mathscr{B}_{uni}(\Omega_0 \to \Omega)$ can be written as $u_c r$ where $r = P_0 r P_0$ is a unitary mapping of $\Omega_0$ on itself: $r^\dagger r = r r^\dagger = P_0$. Introducing $p_c = P_0 u_c = (P_0 P P_0)^{\frac{1}{2}}$, we have

$$p \equiv P_0 u = p_c r \text{ is positive} \Leftrightarrow r = P_0 \,. \tag{5.1}$$

*Proof*: "$\Leftarrow$" "If" is trivial and "$\Rightarrow$" follows by observing that if $p = p_c r$ is positive, it equals the positive square root of $p^2 = (p_c r)(p_c r)^\dagger = p_c (r r^\dagger) p_c = p_c^2$. Q.E.D. In view of this result, Jørgensen's theorem of uniqueness from (2.10b), written as

$$p = P_0 u \text{ is positive} \Leftrightarrow p = p_c \Leftrightarrow u = u_c \tag{5.2}$$

is obvious. It is very simple to modify the proofs of the results in (5.1) and (5.2) so as to obtain the theorem in (2.10a) for the arbitrary $U \in \mathscr{B}_{UNI}(\Omega_0 \to \Omega)$: Writing this as $U = U_c R$ where $R$ is a unitary operator which maps $\Omega_0$ (and therefore also $\Omega_0^\perp$) on itself, one finds that $U_{even}$ is positive iff $R = 1$ - that is iff $U = U_c$.

Consider next Klein's theorem of uniqueness from of (3.16), generalized as follows.

For given $u \in \mathscr{B}_{uni}(\Omega_0 \to \Omega)$, let $|\eta_0\rangle$; $\eta = 1, \ldots, g$, be an orthonormal basis in $\Omega_0$ and let $|\eta\rangle = u|\eta_0\rangle$. Proceeding as in (3.12)-(3.13) one sees that $\varepsilon \geq 0$ defined by

$$\varepsilon(u)^2 = \Sigma_\eta \||\eta\rangle - |\eta_0\rangle\|^2 = \text{Tr}\{2P_0 - p - p^\dagger\}; \quad p = P_0 u = p_c r, \tag{5.3}$$

is independent of how the basis vectors $|\eta_0\rangle$ are chosen and thus is a measure $\varepsilon(u)$ of how much $u$ changes an arbitrary orthonormal basis in $\Omega_0$. Klein showed that

$$\varepsilon(u)^2 \text{ attains minimum} \Leftrightarrow u = u_c \tag{5.4}$$

We shall now see how (5.1) can be used to derive this result in a slightly stronger version: Abbreviating $\varepsilon(+u_c)$ to just $\varepsilon_c$, we have

$$\varepsilon_c^2 = 2\mathrm{Tr}\{P_0 - p_c\} \leq \varepsilon(u)^2 \leq 2\mathrm{Tr}\{P_0 + p_c\} = 4g - \varepsilon_c^2, \qquad (5.5)$$

where minimum / maximum is attained iff $u = u_c / -u_c$.

*Proof*: Evaluate the trace in an orthonormal $\Omega_0$-basis $|\eta_0\rangle$ of eigenfunctions of $p_c$, say $p_c|\eta_0\rangle = p_{c\eta}|\eta_0\rangle$ where $p_{c\eta} > 0$ for all $\eta$. With the orthonormal vectors $|\eta_r\rangle = r|\eta_0\rangle$ satisfying $\||\eta_r\rangle\|^2 = \||\eta_0\rangle\|^2 = 1$ we then get

$$\langle \eta_0 | (p + p^\dagger) | \eta_0 \rangle = p_{c\eta}(\langle \eta_0 | \eta_r \rangle + \langle \eta_r | \eta_0 \rangle) = p_{c\eta}(1 + 1 - \||\eta_r\rangle - |\eta_0\rangle\|^2) \qquad (5.6)$$

Hence, with $\Sigma_\eta p_{c\eta} = \mathrm{Tr}\{p_c\}$,

$$\mathrm{Tr}\{2P_0 - (p + p^\dagger)\} = 2\mathrm{Tr}\{P_0 - p_c\} + \Sigma_\eta p_{c\eta} \||\eta_r\rangle - |\eta_0\rangle\|^2) \qquad (5.7)$$

Minimum / maximum of this quantity is attained iff $r|\eta_0\rangle = |\eta_0\rangle / -|\eta_0\rangle$ for all $\eta$. That is, iff $r = P_0 / -P_0$. The result $4g - \varepsilon_c^2$ for the upper limit in (5.5) is found by writing $P_0 + p_c$ in the lower as $2P_0 - (P_0 - p_c)$ Q.E.D.

We note that since $p_c = \cos\theta P_0$, the lower/upper limit in (5.5) can be written as

$$2\mathrm{Tr}\{P_0 \mp p_c\} = 2\mathrm{Tr}\{(1 \mp \cos\theta)P_0\} = \begin{cases} 4\mathrm{Tr}\{\sin^2\frac{\theta}{2} P_0\} \\ 4\mathrm{Tr}\{\cos^2\frac{\theta}{2} P_0\} \end{cases} \qquad (5.8)$$

*Further elaborations of Klein's theorem of uniqueness*

As derived above, Klein's theorem is a direct consequence of $p \equiv P_0 u$ being positive, that is of the characteristic property in Jørgensen's theorem. This intimate relation between the two theorems is not clear in the following alternative derivation which, however, is of its own interest because it is an almost immediate generalization of the simple geometric proof given in sec. 3. Furthermore it leads to the generalizations of the expressions for the lower/ lower limits found in (3.16) and of the theorem (3.19) for $U_c$. We start with a few preparations.

In section 3, the crucial point in the proof of Klein's theorem is the observation that $|x1\rangle$ in fig. 3.1b is uniquely characterized as the unit vector in $\Omega$ which is nearest to the unit vector $|x_0 1\rangle$ in $\Omega_0$. This result is generalized as follows. By definition, the projection $P|a\rangle$ of a given unit vector $|a\rangle$ is the vector in $\Omega$ which is nearest to $|a\rangle$. The *unit* vector in $\Omega$ which is nearest to $|a\rangle$ is then the *unit* vector along $P|a\rangle$. Although this latter observation is almost self evident, it is verified strictly in appendix B. It means that if $|a\rangle$ is an eigenvector in $\Omega_0$ of $P_0 P P_0$, the nearest unit vector in $\Omega$ is

$$\langle a|P_0 P P_0|a\rangle^{-\frac{1}{2}} P|a\rangle = P(P_0 P P_0)^{-\frac{1}{2}}|a\rangle = U_c|a\rangle \ (= u_c|a\rangle) \tag{5.9}$$

Similarly, the unit vector in $\Omega$, which is most *distant* from $|a\rangle$, is $-U_c|a\rangle$. And analogously: If a unit vector $|b\rangle$ is an eigenvector in $\Omega_0^\perp$ of $Q_0 Q Q_0$, then the unit vector in $\Omega^\perp$, which is nearest to /most distant from $|b\rangle$, is $+U_c|b\rangle / -U_c|b\rangle$.

Klein's theorem in the form from (5.5) now follows directly from the result in (5.9). *Proof*: Choosing the $|\eta_0\rangle$'s in (5.3) as an orthonormal set of eigenvectors for $P_0 P P_0$, the term in the sum with $|\eta\rangle = u|\eta_0\rangle$ satisfy

$$\|(u_c - P_0)|\eta_0\rangle\| \leq \|(u - P_0)|\eta_0\rangle\| \leq \|(u_c + P_0)|\eta_0\rangle\| \tag{5.10}$$

where minimum / maximum is attained iff $u|\eta_0\rangle = +u_c|\eta_0\rangle / -u_c|\eta_0\rangle$. Since this is true for each $|\eta_0\rangle$, the result aimed at follows. *Q.E.D.*

To obtain geometric expressions for the lower and the upper limits in (5.5) we just have to choose the orthonormal eigenvectors $|\eta_0\rangle$ in (5.3) explicitly as an orthonormal basis $|z_0 i\rangle$; $i = 1,2,..$, in $\Omega_0 \cap \Omega$ supplemented by the $|x_0 \mu\rangle$; $\mu = 1,2,..$, from (4.3). For the first of these we obviously have

$$0 \leq \|u|z_0 i\rangle - |z_0 i\rangle\| \leq 2 \tag{5.11}$$

where minimum/maximum is attained iff $u|z_0 i\rangle = +|z_0 i\rangle / -|z_0 i\rangle$. For the second, we can show that

$$d_{\mu-} \equiv 2\sin\frac{\theta_\mu}{2} \le \||u|x_0\mu\rangle - |x_0\mu\rangle\| \le d_{\mu+} \equiv 2\cos\frac{\theta_\mu}{2} \tag{5.12}$$

where minimum/maximum is attained iff $u|x_0\mu\rangle = +u_c|x_0\mu\rangle / -u_c|x_0\mu\rangle$. *Proof*: With $u_c|x_0\mu\rangle = |x\mu\rangle$, the extreme lengths $d_{\mu\mp}$ are found by use of the rearrangement $\||x\mu\rangle \mp |x_0\mu\rangle\|^2 = 2 \mp 2\langle x\mu|x_0\mu\rangle = 2(1 \mp \cos\theta_\mu)$. *Q.E.D.* The simple meanings of the lengths $d_{\mu\mp}$ are illustrated in fig. 4.1. One sees that $d_{\mu-}^2 + d_{\mu+}^2 = 4$ and that

$$\varepsilon_c^2 = \Sigma_\mu d_{\mu-}^2 \tag{5.13}$$

It is simple to check the agreement with $2\text{Tr}\{P_0 - p_c\}$ from (5.8).

Assume next that the whole Hilbert space is of a finite dimension $G$. Then $E(U) > 0$ defined by

$$E(U)^2 = \Sigma_\eta \||\eta\rangle - |\eta_0\rangle\|^2 = 2\text{Tr}\{1 - U_H\} \tag{5.14}$$

measures how much a given $U \in \mathscr{B}_{UNI}(\Omega_0 \to \Omega)$ changes one orthonormal basis $|\eta_0\rangle$; $\eta = 1,2,...,G$, when transforming it to another $|\eta\rangle = U|\eta_0\rangle$. Again this measure is independent of how the $|\eta_0\rangle$-basis is chosen. Letting it be an orthonormal basis $|Z_0 i\rangle$; $i = 1,2,..$, in the space of intersections $\omega^\perp = (\Omega_0 \cap \Omega) \oplus (\Omega_0^\perp \cap \Omega^\perp)$ supplemented by the basis of the $|x_0\mu\rangle$ and $|y_0\mu\rangle$; $\mu = 1,2,..$, in $\omega$, we get

$$E(U)^2 = \Sigma_i \|U|Z_0 i\rangle - |Z_0 i\rangle\|^2 + \Sigma_\mu \|U|x_0\mu\rangle - |x_0\mu\rangle\|^2 + \Sigma_\mu \|U|y_0\mu\rangle - |y_0\mu\rangle\|^2 \tag{5.15}$$

Here the terms in the first sum are subject to the same condition as in (5.11) above (with $U$ instead of $u$), those in the second are similarly subject to the condition (5.12) – and it is easily checked that so are those in the last sum if we replace $|x_0\mu\rangle$ in (5.12) with $|y_0\mu\rangle$. With the same $\varepsilon_c = e(+u_c)$ as in (5.5) we therefore get

$$E(\pm U_c)^2 = 2\text{Tr}\{1 \mp U_{c,even}\} = \begin{cases} 2\varepsilon_c^2 \\ 4G - 2\varepsilon_c^2 \end{cases}. \tag{5.16}$$

And corresponding to the result (5.5): For any $U \in \mathcal{B}_{UNI}(\Omega_0 \to \Omega)$ we have

$$2\varepsilon_c^2 \leq E(U)^2 \leq 4G - 2\varepsilon_c^2 \tag{5.17}$$

where the minimum /maximum is attained iff $U = +U_c / -U_c$. Note that if an orthonormal basis consists of one for $\Omega_0$ and one for $\Omega_0^\perp$ then $U_c$ always changes these two equally.

If the dimension $G$ is infinite, the measure $E(U)$ is not defined for all $U \in \mathcal{B}_{UNI}(\Omega_0 \to \Omega)$. In particular not for $U = -U_c$. It is, however, still defined for *some $U$'s* - an infinite number of terms in the sum for $E(U_c)^2$ in (5.15) vanish and we are left with $E(U_c)^2 = 2\varepsilon_c^2$. This means that the minimum-part of the least square characterization survives in the following form: For all those $U \in \mathcal{B}_{UNI}(\Omega_0 \to \Omega)$ for which $E(U)$ exists one has

$$2\varepsilon_c^2 \leq E(U)^2 \tag{5.18}$$

where the minimum is attained iff $U = U_c$.

# 6: On previous approaches to Klein's theorem of uniqueness

As derived first in sec. 5, Klein's unique least square characteristic of $u_c = Pp_c^{-1}$ is a direct consequence of Jørgensen's: $p = P_0 u$ is positive. In this section we look into how it previously has been such derived that it appeared as a surprising incident that the two characterized the same transformation. As we shall see, our theory in sec. 5 cuts through quite a few pages of previous theory on this theme. We start with an example which shows that while Klein's characteristic has a strong geometric appeal, Jørgensen's is much more easy to use when it comes to reveal that a given $u \in \mathscr{B}_{UNI}(\Omega_0 \to \Omega)$ is identical to $u_c$.

### A: *Jørgensen's theorem applied to Van Vleck's $e^{g_c}$*

Jørgensen [11] found his theorem from (5.2), without yet being aware of Klein's previous paper [9]. He knew that Van Vleck's $\bar{u}_c = e^{g_c} P_0$ is identical to $u_c$ through the explicitly determined first orders in the perturbation expansions and he formulated some general recurrence relations which, interlocking the expansions of $u$ and the effective Hamiltonian $A = u^\dagger H u$, revealed that $e^{g_c} P_0$ must be the same as $u_c$ through *all* orders. Then he found his theorem in (5.2) which is purely geometric in the sense that it has no reference to the perturbation problem which defines the two spaces in the mapping $u: \Omega_0 \to \Omega$. It leads to the identity $e^{g_c} P_0 = u_c$ by the following almost trivial consideration: Since $g_c$ is *odd* and antihermitean,

$$p = P_0 u = P_0 e^{g_c} P_0 = P_0(1 + \tfrac{1}{2!} g_c^2 + \tfrac{1}{4!} g_c^4 + ...)P_0, \tag{6.1}$$

is hermitean – and because of continuity ( $p_c \to P_0$ as $\lambda \to 0$ in $H = H_0 + \lambda V$ ) it is also positive.

Similarly one sees that the *even* part of $e^{g_c}$ is positive. Hence, by Jørgensen's theorem in the form from (2.10a), we have

$$e^{g_c} = U_c = P(P_0 P P_0)^{-\tfrac{1}{2}} + Q(Q_0 Q Q_0)^{-\tfrac{1}{2}}. \tag{6.2}$$

*Pedantic note*: When Jørgensen first pointed this out, he ought to have shown that $\Omega_0^\perp$ and $\Omega^\perp$ are non-orthogonal iff $\Omega_0$ and $\Omega$ are so. In the present paper this latter result is pointed out in the beginning of sec.2.

### B: *Klein's original proof of $u_c$'s unique minimal-effect*

Klein [9] wants to locate the minimum of $\varepsilon^2$ from (5.3) when the $|\eta_0\rangle$'s vary in $\Omega_0$ while the $|\eta\rangle$'s are kept fixed as the orthonormal eigenvectors $|\eta_c\rangle$ of the exact Hamiltonian $H$ in $\Omega$. Writing these as

$$|\eta_c\rangle = u_c |\eta_{c0}\rangle \; ; \; u_c = P p_c^{-1}, \tag{6.3}$$

where the $|\eta_{c0}\rangle$'s are the eigenvectors in $\Omega_0$ of the canonical effective Hamiltonian, he splits $\varepsilon^2$ in a sum of two,

$$\varepsilon^2 = \Sigma_\eta \||\eta_0\rangle - |0\eta_c\rangle\|^2 + \Sigma_\eta \|Q_0|\eta_c\rangle\|^2 \equiv \varepsilon_a^2 + \varepsilon_b^2, \tag{6.4}$$

where the vectors $|0\eta_c\rangle \equiv P_0|\eta_c\rangle$ form a non-ortonormal basis $\Omega_0$ and where $\varepsilon_b^2$ is independent of the variable $|\eta_0\rangle$'s. From a theorem due to Carlson and Keller [14], Klein knows that $\varepsilon_a^2$ attains its unique minimum when the $|\eta_0\rangle$'s are those obtained by applying Løwdin's symmetric orthonormalization method [15,16] to the $|0\eta_c\rangle$. He then takes it for granted (see the text to his eqn. (II.B.2)) that Løwdin's method gives the vectors

$$(P_0 P P_0)^{-\frac{1}{2}} |0\eta_c\rangle = p_c^{-1} |\eta_c\rangle = p_c^{-1} P |\eta_c\rangle = u_c^\dagger |\eta_c\rangle, \tag{6.5}$$

which we recognize as the orthonormal eigenvectors of the canonical effective Hamiltonian $A_c = u_c^\dagger H u_c$. This is correct, but some extra arguments are needed to make sure that (6.5) actually represents a use of Löwdin's method. Reformulating this method as shown in appendix E, it can be seen simply by noting that $B$ from (E2) in the present case can be written as

$$B = \Sigma_\eta |0\eta_c\rangle\langle 0\eta_c| = P_0(\Sigma_{\eta'}|\eta_c\rangle\langle \eta_c|)P_0 = P_0 P P_0 \tag{6.6}$$

In this way Klein has proven his theorem of uniqueness: $\varepsilon_a^2$ (and thereby $\varepsilon^2$) in (6.4) attains minimum iff $|\eta_0\rangle$ equals $|\eta_{c0}\rangle$ from (6.3) for all $\eta$. Here we can add the stronger result that

$$\text{Tr}\{(P_0 - p_c)^2\} \leq \varepsilon_a^2(u) = \varepsilon^2(u) - \varepsilon_b^2 \leq \text{Tr}\{(P_0 + p_c)^2\} \tag{6.7}$$

where minimum / maximum is attained iff $u = +u_c / -u_c$. The limits follow directly from (6.4) by insertion of

$$\varepsilon_b^2 = \Sigma_\eta \langle \eta_c | Q_0 | \eta_c \rangle = \text{Tr}\{u_c^\dagger (1 - P_0) u_c\} = \text{Tr}\{P_0 - p_c^2\}. \tag{6.8}$$

<p align="center">C: <i>Comments on Klein's original approach</i></p>

Instead of measuring the *difference* between the two orthonormal sets by $\varepsilon^2$, Klein actually measures their *similarity* defined as $\xi \equiv 1 - g^{-1}\varepsilon^2$ (which equals 1 (=100%) iff $|\eta_0\rangle = |\eta\rangle$ for all $\eta = 1, 2, ..., g$). He can then assert that the canonical effective Hamiltonian is uniquely determined as the one whose eigenvectors bear maximum similarity to those of the real *H*. That is of course why the characteristic is so appealing – but mathematically, the formulation conceals the fact that $\varepsilon(u)$ is independent of the $\Omega_0$-basis which *u* transforms to a basis $\Omega$. It only depends on the transformation itself. This observation is crucial for the calculations in the present paper.

Klein attempted to show that Van Vleck's $\bar{u}_c = e^{g_c} P_0$ equals the canonical $u_c = P p_c^{-1}$ by proving that $\varepsilon_a^2$ in (6.4) attains minimum when the variable $|\eta_0\rangle$ equals $|\bar{\eta}_{c0}\rangle \equiv e^{-g_c}|\eta_c\rangle$ for all $\eta$. Unfortunately, as pointed out later by Jørgensen [17], his proof failed on some rather treacherous details. However, the minimal-effect could instead be verified in two steps by first using Jørgensen's theorem in (5.2) to show that $\bar{u}_c$ is the same as $u_c$. To quote Brandow [12 p. 220-221]: " This Hermiticity condition (together with the positive definiteness) is undoubtedly the most general and straightforward way to formulate the desired minimal effect". Brandow could, however, still have good reasons for not just forgetting about a verification by use of Klein's original direct approach. See the next subsection.

D: *Brandow's correction of Klein's variation calculation*

Klein's original approach to his unique characteristic of $u_c$ did not –as our proof of (5.5) in the previous section - reveal that it is a direct consequence of Jørgensen's $p = P_0 u$ being positive. Furthermore, Kleins proof was built on non-trivial theory developed in other contexts, first by Löwdin [15,16] and then by Carlson and Keller [14]. This is probably a reason why Brandow [12] found it desirable to circumvent these layers of earlier theory in a fairly elaborate appendix reserved a correct version of Klein's direct variation calculation. Starting as in (6.4) above, Brandow wrote the variable $|\eta_0\rangle$ as $e^{-F}|\eta_c\rangle$ where $F = -F^\dagger$ is variable and he then proved that

$$\varepsilon_a^2 = \Sigma_\eta \left\| (e^{-F} - P_0) e^{g_c} |\bar{\eta}_{c0}\rangle \right\|^2 \tag{6.9}$$

attains minimum if $|\eta_0\rangle = e^{-F}|\eta_c\rangle$ equals $|\bar{\eta}_{c0}\rangle \equiv e^{-g_c}|\eta_c\rangle$ for all $\eta$. His calculations can be substantially abbreviated by first rewriting to

$$\varepsilon_a^2 = \Sigma_\eta \left\| (Y - \bar{p}_c) |\bar{\eta}_{c0}\rangle \right\|^2$$
$$= \text{Tr}\{P_0 + \bar{p}_c^2 - \bar{p}_c(Y + Y^\dagger)\} = \text{Tr}\{(P_0 - \bar{p}_c)^2 - \bar{p}_c(Y + Y^\dagger - 2)\} \tag{6.10}$$

where $Y = e^{-F} e^{g_c}$ and where $\bar{p}_c = P_0 e^{g_c} P_0$ is positive (by the same argument as in (6.1) above). When $F$ varies around $g_c$, $Y$ varies around unity and can therefore be written as $Y = 1 + \delta Y$ where the infinitesimal part $\delta Y$ is antihermitean. Hence $Y + Y^\dagger - 2$ in (6.10) vanishes for small variations $F = g_c + \delta F$ and $\varepsilon_a^2$ is therefore stationary at $F = g_c$. Brandow reaches this conclusion after a procedure which amount to finding $\delta Y$ explicitly in terms of $\delta F$ by use of some not quite simple Hausedorff-expansions of the exponentials.

Brandow ends by concluding that $\varepsilon_a^2$ as a function of $F$ has a stationary point at $F = g_c$ which, he argues, , must be the unique global minimum for reasons of continuity. We make the pedantic note that it is $u = e^F P_0$ which is uniquely determined at minimum – thus $e^F P_0$ equals $e^{g_c} P_0$ for any of the many antihermitean $F$ for which solves $e^F = e^{g_c} P_0 + U Q_0$ where $U \in \mathscr{B}_{UNI}(\Omega_0 \to \Omega)$.

*A formulation in terms of a norm for operators*

We end this review-like section by a formulation of Klein's characteristic given by Jørgensen [18]. The proof is simplified and the approach is used for two further characteristics of the canonical $u_c$. When the space $\Omega_0$ is of finite dimension, the set of all operators of the form $v = vP_0$ can be considered as a space of vectors with inner product and norm defined by [19 p. 132]

$$(v_1, v_2) = \mathrm{Tr}\{v_1^\dagger v_2\} \quad \text{and} \quad ((v))^2 = (v, v) = \mathrm{Tr}\{v^\dagger v\}. \tag{6.11}$$

Doing so, it can be shown (see below) that for any $u \in \mathscr{B}_{uni}(\Omega_0 \to \Omega)$,

$$2\mathrm{Tr}\{P_0 - p_c\} \leq ((u - P_0))^2 \leq 2\mathrm{Tr}\{P_0 + p_c\}, \tag{6.12a}$$

$$\mathrm{Tr}\{(P_0 - p_c^{-1})^2\} \leq ((u - \mathscr{U}))^2 \leq \mathrm{Tr}\{(P_0 + p_c^{-1})^2\}, \tag{6.12b}$$

$$\mathrm{Tr}\{(P_0 - p_c)^2\} \leq ((u - PP_0))^2 \leq \mathrm{Tr}\{(P_0 + p_c)^2\}, \tag{6.12c}$$

where, in all three cases, minimum / maximum is attained iff $u = u_c / -u_c$.

Note how the minimum-results can be formulated in an intuitively descriptive way. The second, for instance, asserts that the canonical $u_c$ is the $u \in \mathscr{B}_{uni}(\Omega_0 \to \Omega)$ which comes nearest to Bloch's $\mathscr{U}$. Note also that $((u - P_0))^2$ in the (6.12a) equals $\varepsilon(u)^2$ from (5.3). Hence the first result is the same as Klein's result in the form from (5.5).

We shall now see that all three results (6.12) follow almost directly from the following well known rule, fundamental for any Hilbert space of vectors $x, y, \ldots$, with inner product $(x, y)$ and norm $\|x\|^2 = (x, x)$: If $\|x\|^2 = \|y\|^2$ then

$$-2\|x\|^2 \leq 2\,\mathrm{Re}(x, y) = (x, y) + (y, x) \leq 2\|x\|^2 \tag{6.13}$$

where minimum / maximum is attained iff $x$ equals $y / -y$.

*Proof*: With $((u))^2 = \text{Tr}(u^\dagger u) = \text{Tr}(P_0)$, $((\mathscr{V}))^2 = \text{Tr}\{p_c^{-2}\}$ and $((PP_0))^2 = \text{Tr}\{p_c^2\}$ we get

$$((u - P_0))^2 = \text{Tr}\{P_0 + P_0\} - 2\,\text{Re}(P_0, u), \tag{6.14a}$$

$$((u - \mathscr{V}))^2 = \text{Tr}\{P_0 + p_c^{-2}\} - 2\,\text{Re}(\mathscr{V}, u), \tag{6.14b}$$

$$((u - PP_0))^2 = \text{Tr}\{P_0 + p_c^2\} - 2\,\text{Re}(PP_0, u), \tag{6.14c}$$

where $u = u_c r = P p_c^{-1} r$ as in (5.1). In (6.14a), $(P_0, u)$ equals $\text{Tr}(P_0 u) = \text{Tr}\{p_c r\} = (x, y)$ where $x = p_c^{\frac{1}{2}}$ and $y = xr$ satisfy $\|x\|^2 = \|y\|^2 = \text{Tr}\{p_c\}$. The rule (6.13) then asserts that $-2\,\text{Re}(P_0, u)$ attains minimum $\text{Tr}\{-2p_c\}$ / maximum $\text{Tr}\{+2p_c\}$ iff $r = P_0 / - P_0$. This is the result aimed at. The remaining results are shown similarly by writing $(\mathscr{V}, u) = \text{Tr}\{(p_c^{-2} P \cdot P p_c^{-1} r\}$ as $\text{Tr}\{p_c^{-\frac{1}{2}} \cdot p_c^{-\frac{1}{2}} r\}$ and $(PP_0, u) = \text{Tr}\{(P_0 P \cdot P p_c^{-1} r\}$ as $\text{Tr}\{p_c^{\frac{1}{2}} \cdot p_c^{\frac{1}{2}} r\}$. Q.E.D.

## A: Some definitions and lemmas

Obviously the arbitrary operator $B$ can be decomposed into the sum $B_H + B_A$ of its hermitean and its anti-hermitean part,

$$B_H = \tfrac{1}{2}(B+B^\dagger) \quad \text{and} \quad B_A = \tfrac{1}{2}(B-B^\dagger). \tag{A1}$$

By definition [20 p. 313, 21 p. 257] an operator $B$ is *positive* if $\langle a|B|a\rangle$ is a non-negative real number for any non-zero vector $|a\rangle$. A positive operator is thus same as a hermitean operator with no negative eigenvalues. Any operator of the form $B^\dagger B$ or $BB^\dagger$ is positive. If $B$ is positive, the equation $X^2 = B$ has one, and only one, positive solution $X$, namely $X = B^{\frac{1}{2}}$. To distinguish this from other solutions, it is sometimes called *the positive* square root.

In the mathematical literature, an *even* operator (defined as in (1.7)) is said to be *reduced* by the subspace $\Omega_0$ [22 p. 275, 23 p. 571, 24 p. 146]. One sees that $B$ is reduced by $\Omega_0$ iff $B|a\rangle$ and $B^\dagger|a\rangle$ both belong to $\Omega_0$ when $|a\rangle$ does so. One can of course define the properties *even* and *odd* relative to *any* given subspace in the Hilbert space. In the present paper, unless something else is asserted explicitly, the designation always refers to the subspace $\Omega_0$. The operator $R_0 = P_0 - Q_0$ from (1.8) is called the reflection in $\Omega_0$ because it leaves vectors in $\Omega_0$ unchanged while it changes the sign of vectors orthogonal to $\Omega_0$. One can of course define a reflection operator relative to *any* given subspace in the Hilbert space. Thus $R = P - Q$ is the reflection in $\Omega$. One sees that an operator $B$ is a reflection iff it is unitary as well as hermitean. We disregard the extreme cases $B = 1$ and $B = -1$.

Let now $P_\omega$ be the projector on a certain subspace $\omega$ and let $c_4$ denote an antihermitean operator which is unitary within $\omega$ (i.e., $c_4 c_4^\dagger = c_4^\dagger c_4 = P_\omega$) and fulfils $c_4 = -c_4^\dagger = P_\omega c_4 P_\omega$ so that

$$c_4^1 = c_4, \quad c_4^2 = -P_\omega, \quad c_4^3 = -c_4, \quad c_4^4 = P_\omega, \quad c_4^5 = c_4, \; .... \; , \tag{A2}$$

We call such an operator a *quarter turn within* $\omega$. The index "4" is used to remind us that the operators $(P_\omega, c_4, c_4^2, c_4^3)$ form a representation of the group $C_4$. Note that

$$c_4 \text{ is a quarter turn within } \omega \Leftrightarrow ic_4 \text{ is a reflection within } \omega, \tag{A3}$$

The latter assertion means that $ic_4$ can be written as $p - q$ where $p$ and $q$ are orthogonal projection operators satisfying $p + q = P_\omega$.

The results (A2) reminds us of the results $i^1 = i$, $i^2 = -1$, $i^3 = -i$, $i^4 = 1$,.. for the imaginary unit, used to verify the identity $e^{i\varphi} = \cos\varphi + i\sin\varphi$ by expansion of the functions of the angle $\varphi$. With $c_4$ instead of "$i$" in the expansion of the exponential, use of (A2) gives

$$e^{c_4\varphi} = 1 + (-\frac{\varphi^2}{2!} + \frac{\varphi^4}{4!} - ...)P_\omega + c_4(\varphi - \frac{\varphi^3}{3!} + ...) \tag{A4}$$

which, with $Q_\omega = 1 - P_\omega$, can be written as in

$$D(\varphi) \equiv e^{c_4\varphi} = \begin{cases} Q_\omega + P_\omega(\cos\varphi + c_4\sin\varphi) \\ Q_\omega + P_\omega e^{c_4\varphi} \end{cases} \tag{A5}$$

Since $c_4$ is antihermitean, $D(\varphi)$ is unitary. We note that it represents the rotation $C(\varphi)$ of angle $\varphi$ in the group $C_\infty$. In a more loose sense we can think of the orthonormal complement $\omega^\perp$, which is left invariant by $D(\varphi)$, as the axis for that rotation. And we can think of the reflection $R_\omega = P_\omega - Q_\omega$ as representing the reflection $\sigma_h$ in the group $C_{\infty h}$.

## B: Non-orthogonal subspaces

Bloch [7] assumed from a start that the zeroth order eigenspace $\Omega_0$ and the space $\Omega$ of the perturbed eigenfunctions are of the same finite dimension $g$. Defining that $\Omega$ is non-orthogonal to $\Omega_0$ if no vector in $\Omega$ is orthogonal to all vectors in $\Omega_0$, he found that this definition is symmetric in the sense that $\Omega$ is non-orthogonal to $\Omega_0$ iff $\Omega_0$ is non-orthogonal to $\Omega$. In the present paper we start directly with the symmetric definition: Unless something else is stated explicitly, $\Omega$ and $\Omega_0$ are assumed to be non-orthogonal in the sense that no vector in one of the spaces is orthogonal to all vectors in the other. This definition implies that $\dim\Omega_0 = \dim\Omega$. So, if this dimension is finite, the two ways of starting are equivalent. However, a simple example shows that this is not so in general: Assume that $\Omega_0$ is spanned by some orthonormal vectors $|1\rangle, |2\rangle, |3\rangle,...$ and $\Omega$ by the subset $|2\rangle, |4\rangle, |6\rangle,...$ . Then no vector in $\Omega$ is orthogonal to all vectors in $\Omega_0$. But since each of the vectors $|1\rangle, |3\rangle,...$ in $\Omega_0$ is orthogonal to all vectors in $\Omega$, the two subspaces are *not* non-orthogonal by the symmetric definition we now use.

For any given unit vector $|e_0\rangle \in \Omega_0$, the projection $P|e_0\rangle$ is non-zero and it is the unique vector in $\Omega$ which is nearest to $|e_0\rangle$. The nearest *unit* vector in $\Omega$ is

$$|e_n\rangle \equiv \||P|e_0\rangle\|^{-1} P|e_0\rangle = \langle e_0|P_0 P P_0|e_0\rangle^{-\frac{1}{2}} P|e_0\rangle \tag{B1}$$

(index *n* for "nearest"). This intuitively rather obvious assertion can be verified strictly by proving the following more carefully stated result: Let $|e'\rangle$ be an arbitrary unit vector in $\Omega$. Then

$$\||e_n\rangle - |e_0\rangle\|^2 \leq \||e'\rangle - |e_0\rangle\|^2 = \||e'\rangle - P|e_0\rangle\|^2 + \|Q|e_0\rangle\|^2 \leq \||e_n\rangle + |e_0\rangle\|^2 \tag{B2}$$

where the minimum /maximum is given by

$$\||e_n\rangle \mp |e_0\rangle\|^2 = 2 \mp 2\langle e_0|P_0 P P_0|e_0\rangle^{+\frac{1}{2}} \quad (= 2 \mp \||P|e_0\rangle\|) \tag{B3}$$

and is attained iff $|e'\rangle = +|e_n\rangle / -|e_n\rangle$.

*Proof*: The sign of equality at the centre of (B2) is verified by writing $|e'\rangle - |e_0\rangle$ as $P(|e_n\rangle - |e_0\rangle) - Q|e_0\rangle$. The term $\||e'\rangle - P|e_0\rangle\|^2$ attains minimum / maximum when $|e'\rangle$ points along $+P|e_0\rangle / -P|e_0\rangle$. The extreme-values in (B3) follow directly from $\langle e_n|e_0\rangle = \langle e_0|e_n\rangle = \langle e_0|P|e_0\rangle \langle e_0|P|e_0\rangle^{-\frac{1}{2}} = \langle e_0|P|e_0\rangle^{+\frac{1}{2}}$ Q.E.D.

*Lemma*: With $u_c = P(P_0 P P_0)^{-\frac{1}{2}}$ from (2.8), $|e_n\rangle$ equals $|e_c\rangle \equiv u_c|e_0\rangle$ if and only if $|e_0\rangle$ is an eigenvector of $P_0 P P_0$.

*Proof*: Since the spaces $\Omega_0$ and $\Omega$ are non-orthogonal, the difference

$$|\delta\rangle = |e_c\rangle - |e_n\rangle = P|\delta_0\rangle; \quad |\delta_0\rangle = (P_0 P P_0)^{-\frac{1}{2}}|e_0\rangle - \langle e_0|P_0 P P_0|e_0\rangle^{-\frac{1}{2}}|e_0\rangle, \quad (B4)$$

vanishes iff $|\delta_0\rangle = 0$. Q.E.D.

## C: Normal operators and polar decomposition

Any given operator $B$ has a unique decomposition $B = B_H + B_A$ into the sum of a hermitean part $B_H = (B + B^\dagger)/2$ and an antihermitean part $B_A = (B - B^\dagger)/2$. If

$$[B_H, B_A] = 0 \ (\Leftrightarrow [B^\dagger, B] = 0), \quad (C1)$$

the operator $B$ is called *normal* [20 p. 298, 22 p. 269, 24 p. 98]. Thus, if $B$ is hermitean, antihermitean or unitary, it is also normal. A normal operator $B$ has a unique spectral resolution

$$B = \Sigma_\alpha z_\alpha p_\alpha \ ; \quad z_\alpha = x_\alpha + i y_\alpha, \quad (C2)$$

where $x_\alpha$ and $y_\alpha$ are the real eigenvalues of the commuting hermitean operators $B_H$ and $-i B_A$, where $\alpha \neq \beta \Leftrightarrow z_\alpha \neq z_\beta$ and where $p_\alpha$; $\alpha = 1, 2, ...$, are the projectors on orthogonal subspaces $\omega_\alpha$; $\alpha = 1, 2, ...$, such that $\Sigma_\alpha p_\alpha = 1$. For given complex function $f(z)$ of the complex number $z$ one defines the function

$$f(B) = \Sigma_\alpha f(z_\alpha) p_\alpha \quad (C3)$$

of the normal operator $B$ from (C2). Thus $f(B)$ commutes with $B$ and is normal. Note: If $z_\alpha$ is zero for a certain $\alpha$ in the sum (C2), the term $f(z_\alpha) p_\alpha$ can only be excluded from the sum (C3) if $f(0) = 0$. A special function is

$$|B|^r = \Sigma_\alpha |z_\alpha|^r p_\alpha; \quad |z_\alpha| = (|x_\alpha|^2 + |y_\alpha|^2)^{\frac{1}{2}}, \quad (C4)$$

defined for any positive real number *r*. In particular,

$$|B|^2 = B^\dagger B = B B^\dagger = |B_H|^2 + |B_A|^2 \quad (C5)$$

($= B_H^2 - B_A^2$). If $B$ has an inverse, so has $|B|$ and (C4) holds for *all* real *r*. The inverse can be written either as $|B|^{-2} B^\dagger$ or as $B^\dagger |B|^{-2}$ and since these two are identical we can abbreviate to just

$$B^{-1} = \frac{B^\dagger}{|B|^2} = \frac{B_H - B_A}{|B|^2} \quad (= \Sigma_\alpha z_\alpha^{-1} p_\alpha \,; \quad z_\alpha^{-1} = \frac{z_\alpha^*}{|z_\alpha|^2}). \tag{C6}$$

With the same abbreviated notation, the invertible normal operator $B$ can be given by its so-called *polar decomposition* [20 p. 315, 24 p. 242],

$$B = |B|U = U|B| \quad \text{where} \quad U = \frac{B}{|B|} = \Sigma_\alpha \frac{z_\alpha}{|z_\alpha|} p_\alpha \tag{C7}$$

is seen to be unitary ( satisfying $UU^\dagger = U^\dagger U = 1$). There are no other ways whereby the invertible normal $B$ can be written as a product of unitary and a positive operator. *Proof*: If $|B|U = RE$ where $R$ is positive and $E$ unitary then $|B|UU^\dagger|B| = REE^\dagger R \Leftrightarrow |B|^2 = R^2 \Leftrightarrow |B| = R$. Similarly if $U|B| = ER$. Q.E.D. For the operators in the polar decomposition (C7) we see that

if $B = B_H$ then $U = U^\dagger$, i.e. $U$ is a *reflection*,

if $B = B_A$ then $U = -U^\dagger$, i.e. $U$ is a *quarter turn*, \qquad (C8)

(the latter defined as in (A2)). As already mentioned, a unitary operator $U$ is always normal. Writing it as in (C2) with $z_\alpha = x_\alpha + iy_\alpha = e^{i\chi_\alpha}$ we get

$$U = e^{i\chi} = \cos\chi + i\sin\chi \quad \text{where} \quad \chi = \Sigma_\alpha \chi_\alpha p_\alpha. \tag{C9}$$

Each angle $\chi_\alpha$ is uniquely determined by $U$ except for a trivial multiple of $2\pi$. To make the value unique, we demand that $-\pi < \chi_\alpha \leq \pi$. With

$$\chi_{odd} = \Sigma_\alpha \frac{\chi_\alpha}{2}(p_\alpha - R_0 p_\alpha R_0) \quad \text{and} \quad \chi_{even} = \Sigma_\alpha \frac{\chi_\alpha}{2}(p_\alpha + R_0 p_\alpha R_0) \tag{C10}$$

we see that $\chi_{even} = 0$ if and only if each $R_0 p_\alpha R_0$ equals a certain $p_{\overline{\alpha}}$ of the projectors $p_1, p_2,...$, characterized by $\chi p_{\overline{\alpha}} = -\chi_\alpha$. We now use these results to understand the warning at the end of sec. 3.

In the simple 3-dimensional case considered there we introduce the two orthonormal vectors in $\omega$ defined by

$$|1_0 \pm\rangle = 2^{-\frac{1}{2}}(|x_0 1\rangle \mp i|y_0 1\rangle) \quad (\Leftrightarrow \begin{cases} |x_0 1\rangle = 2^{-\frac{1}{2}}(|1_0 +\rangle + |1_0 -\rangle) \\ |y_0 1\rangle = 2^{-\frac{1}{2}}i(|1_0 +\rangle - |1_0 -\rangle) \end{cases}) \tag{C11}$$

and $c_4|1_0 \pm\rangle = \pm i|1_0 \pm\rangle$. With $\theta|1_0 \pm\rangle = \theta_1|1_0 \pm\rangle$ they only differ from

$$|1\pm\rangle \equiv 2^{-\frac{1}{2}}(|x1\rangle \mp i|y1\rangle) = e^{c_4\theta}|1_0 \pm\rangle = e^{\pm i\theta_1}|1_0 \pm\rangle \tag{C12}$$

by a phase factor. Because of the imaginary unit, they cannot be drawn in fig. 3.1. However, with $-ic_4 = |1_0 +\rangle\langle 1_0 +| - |1_0 -\rangle\langle 1_0 -|$ we see that the unique solution to $U_c = e^{c_4\theta} = e^{i\chi}$ is

$$\chi_c = -ic_4\theta = 0 \cdot Q_\omega + |1_0 +\rangle\theta_1\langle 1_0 +| - |1_0 -\rangle\theta_1\langle 1_0 -| \tag{C13}$$

with the eigenvalues $0, -\theta_1$ and $+\theta_1$. We also see that $\theta$ equals $|\chi_c|$ and that the complete solution is obtained from $\chi_c$ by adding

$$\begin{aligned}\Delta\chi &= 2\pi(n_0 \cdot Q_\omega + |1_0 +\rangle n_+\langle 1_0 +| + |1_0 -\rangle n_-\langle 1_0 -|) \\ &= c_4(n_+ - n_-)\pi + (n_+ + n_-)\pi P_\omega + 2\pi n_0 \cdot Q_\omega\end{aligned} \tag{C14}$$

where $n_0$, $n_+$ and $n_-$ are arbitrary integers. Since $P_\omega$ and $Q_\omega$ are *even*, $\Delta\chi$ is *odd* iff $n_0 = 0$ and $n_+ = -n_-$. Note also that $i(\chi_c + \Delta\chi)$ can be *odd* without being identical to the canonical $g_c = i\chi_c$.

## D: Properties of the operator *S* defined by two arbitrary subspaces

Let $P$ and $P_0$ project on certain arbitrarily chosen, *not necessarily non-orthogonal*, subspaces $\Omega$ and $\Omega_0$ so that $Q = 1 - P$ and $Q_0 = 1 - P_0$ project on their orthogonal complements. And let *even* and *odd* relative to $\Omega_0$ be defined as in (1.7). Then the operator $S = PP_0 + QQ_0$ is *normal, i.e.*

$$SS^\dagger = PP_0P + QQ_0Q \text{ equals } S^\dagger S = P_0PP_0 + Q_0QQ_0 \,. \tag{D1}$$

*Proof*: Using such results as $P_0QQ_0 = P_0(1-P)Q_0 = -P_0PQ_0$, we first show that $SS^\dagger$ is *even*: $Q_0SS^\dagger P_0$ vanishes because so does the adjoint $P_0SS^\dagger Q_0 = P_0PP_0PQ_0 + P_0QQ_0QQ_0$ where the second term can be written as $-P_0PQ_0QQ_0 = +P_0PP_0QQ_0 = -P_0PP_0PQ_0$. The proof is therefore completed if we can show that $P_0SS^\dagger P_0 = P_0PP_0$ and $Q_0SS^\dagger Q_0 = Q_0QQ_0$. The first of these identities follows by writing the second term in $P_0PP_0PP_0 + P_0QQ_0QP_0$ as $P_0P(1-P_0)PP_0$. The second identity follows analogously. *Q.E.D.*

Appendix C collects some properties of normal operators. Since the operators $S$, $P_0$, $Q_0$, $P$ and $Q$ all commute with

$$|S|^2 = SS^\dagger = S^\dagger S = |S_H|^2 + |S_A|^2 = S_{even} \tag{D2}$$

they also all commute with the positive square root $|S|$. Straightforward rearrangements lead to the identities

$$S_A = S_{odd} = \begin{cases} PP_0 - P_0 P \\ QQ_0 - Q_0 Q \end{cases} \quad \text{and} \quad S_H = S_{even} = \begin{cases} P_0 P + QQ_0 \\ PP_0 + Q_0 Q \end{cases}. \tag{D3}$$

so that, by (D1),

$$[S^\dagger, S] = 2[S_H, S_A] = 2[S_{even}, S_{odd}] = 0. \tag{D4}$$

We note that an interchange $\Omega \leftrightarrow \Omega_0$ of the defining spaces amounts to $(P,Q) \leftrightarrow (P_0, Q_0)$ which replaces $S$ with $S^\dagger$, leaves $S_H = S_{even}$ invariant and changes the sign of $S_A = S_{odd}$.

Insertion of $S_H = S_{even} = |S|^2$ and $|S_A| = |S_{odd}|$ in (D2) gives

$$|S_{odd}| = |S| \cdot (1 - |S|^2)^{\frac{1}{2}} \tag{D5}$$

Since this operator cannot have a negative eigenvalue, the arbitrary eigenvalue $s_\mu$; $\mu = 1,2,...$, of $|S|$ must be in the interval $[0,1]$. Since one and only one angle $\theta_\mu \in [0, \frac{\pi}{2}]$ solves $\cos\theta_\mu = s_\mu$, the identity

$$\cos\theta = |S| \quad (\Leftrightarrow \sin\theta = (1 - \cos^2\theta)^{\frac{1}{2}} = (1 - |S|^2)^{\frac{1}{2}}) \tag{D6}$$

uniquely defines a positive operator $\theta$ with these eigenvalues $\theta_\mu$. By (D5) the defining equation can also be written as

$$|S_{odd}| = \cos\theta \sin\theta = \tfrac{1}{2}\sin 2\theta \tag{D7}$$

As a function of $|S|$, $\theta$ commutes with all operators which commute with $|S|$. Thus $\theta$ commutes with $S$, $S_{even} = S_H$, $S_{odd} = S_A$, $P_0$ and $P$.

# E: The Carlson-Keller theorem for Löwdin's symmetric orthonormalization

Let $P_0$ be the projector on a subspace $\Omega_0$ spanned by a set $|\alpha_0\rangle$; $\alpha = 1,...,g$, of linearly independent vectors, let $|\alpha\rangle$; $\alpha = 1,...,g$, be a set of orthonormal vectors, (not necessarily in $\Omega_0$). How should the latter be defined in terms of the former to give the smallest possible value of

$$\varepsilon^2 \equiv \Sigma_\alpha \||\alpha\rangle - |\alpha_0\rangle\|^2 \tag{E1}$$

To answer this question we first note that in the non-orthonormal basis of the $|\alpha_0\rangle$'s, the operator

$$B = \Sigma_\alpha |\alpha_0\rangle\langle\alpha_0| \tag{E2}$$

is represented by the Gram matrix $G_{\beta\alpha} = \langle\beta_0|\alpha_0\rangle$ - because, as one sees directly

$$B|\alpha_0\rangle = \Sigma_\beta |\beta_0\rangle G_{\beta\alpha} \tag{E3}$$

for all $\alpha$. Considered as an operator on $\Omega_0$, $B$ is invertible and positive (because $\langle x|B|x\rangle > 0$ for any non-zero $|x\rangle \in \Omega_0$), and we can therefore define a positive operator $B^r$ on $\Omega_0$ for any real value of $r$. Since this is represented by the positive matrix $G^r_{\beta\alpha}$, the vectors

$$|\alpha_s\rangle = B^{-\frac{1}{2}}|\alpha_0\rangle = \Sigma_\beta |\beta_0\rangle G^{-\frac{1}{2}}_{\beta\alpha}, \tag{E4}$$

easily checked to be orthonormal, are (as also pointed out in [13 p. 401]) identical to those resulting from Löwdin's method. In terms of these one has

$$\Sigma_\alpha \||\alpha_s\rangle - |\alpha_0\rangle\|^2 \leq \Sigma_\alpha \||\alpha\rangle - |\alpha_0\rangle\|^2 \leq \||\alpha_s\rangle + |\alpha_0\rangle\| \tag{E5}$$

where minimum/maximum is attained iff $|\alpha\rangle = +|\alpha_s\rangle / -|\alpha_s\rangle$ for all $\alpha$.

*Proof*: Introducing the operator $u = \Sigma_\alpha |\alpha\rangle\langle\alpha_s|$ one has

$$\varepsilon^2 = \Sigma_\alpha \|(u - B^{-\frac{1}{2}})|\alpha_s\rangle\|^2 = \text{Tr}\{(u^\dagger - B^{-\frac{1}{2}})(u - B^{-\frac{1}{2}})\} \tag{E6}$$

Since the trace is independent of the basis used for its evaluation, we can replace the $|\alpha_s\rangle$'s with an orthonormal set of eigenvectors $|\alpha_B\rangle$ for $B$ so that we get

$$\varepsilon^2 = \Sigma_\alpha \|u|\alpha_B\rangle - c_\alpha|\alpha_B\rangle\|^2 \text{ where } c_\alpha > 0 \text{ for all } \alpha. \tag{E7}$$

Here the vector difference $u|\alpha_B\rangle - c_\alpha|\alpha_B\rangle$ attains its minimal length iff the two vectors point in the same direction. That is, iff $u|\alpha_B\rangle = |\alpha_B\rangle$. Since this is true for each $\alpha$, $\varepsilon^2$ attains minimum if, and only if $u = P_0$; that is if, and only if $|\alpha\rangle = |\alpha_s\rangle$ for all $\alpha$. With the analogue result for maximal length of $u|\alpha_B\rangle - c_\alpha|\alpha_B\rangle$, the result aimed at follows. Q.E.D.

## F: How much does $u_c$ change the arbitrarily chosen vector in $\Omega_0$

Klein's least square characterization only works when the dimension $g$ of $\Omega_0$ is finite. We do, however, still expect that $u_c$ gives a rather small difference $|\Delta\eta\rangle = u_c|\eta_0\rangle - |\eta_0\rangle$ for any $|\eta_0\rangle \in \Omega_0$. The following considerations, valid also for $g = \infty$, substantiate that expectation. We ask: How much does the transformed $|e_c\rangle = u_c|e_0\rangle \in \Omega$ of an arbitrarily chosen unit vector $|e_0\rangle \in \Omega_0$ deviate from the unique unit vector $|e_n\rangle \in \Omega$ which is nearest to $|e_0\rangle$ (index $n$ for "nearest"). As shown in appendix B, $|e_n\rangle$ is the unit vector which points along $P|e_0\rangle$ (identical to $u_c|e_0\rangle$ iff $|e_0\rangle$ is an eigenvector of $P_0 P P_0$). We shall now make it explicit that length $\delta$ of $|\delta\rangle = |e_n\rangle - |e_c\rangle$ is small even when compared to the length $\Delta$ of $|\Delta\rangle = |e_n\rangle - |e_0\rangle$.

Simple rearrangements give

$$\Delta^2 = \langle\Delta|\Delta\rangle = 2 - 2\langle e_0|P_0 P P_0|e_0\rangle^{+\frac{1}{2}} \quad ,$$

$$\delta^2 = \langle\delta_0|P_0 P P_0|\delta_0\rangle = 2 - 2\langle e_0|P_0 P P_0|e_0\rangle^{-\frac{1}{2}}\langle e_0|(P_0 P P_0)^{\frac{1}{2}}|e_0\rangle. \tag{F1}$$

For evaluation of these quantities, introduce the mean value $\langle B \rangle \equiv \langle e_0|B|e_0\rangle$ of the arbitrary operator $B$, note such results as $\langle P_0 P P_0 \rangle = \langle \cos^2 \theta \rangle$ and use familiar expansions like $\cos x = 1 - \frac{1}{2!}x^2 + \frac{1}{4!}x^4 - ..$ and $(1+x)^n = 1 + nx + \frac{1}{2!}n(n-1)x^2 + ...$ . To keep track of the orders in the expansions we replace $\theta$ with $\lambda\theta$ where $\lambda \in [0,1]$ is an interpolation parameter so that the actual $u_c$ can be considered as the limit $\lambda \to 1$ of $u_c(\lambda) = D(\lambda\theta)$. Straightforward calculation then gives

$$\tfrac{\Delta^2}{2} = 1 - \langle \cos^2 \lambda\theta \rangle^{+\frac{1}{2}} = \tfrac{\lambda^2}{2} <\theta^2> + \lambda^4 + ...,$$

$$\tfrac{\delta^2}{2} = 1 - \langle \cos^2 \lambda\theta \rangle^{+\frac{1}{2}} \langle \cos\lambda\theta \rangle = \tfrac{\lambda^4}{8}(<\theta^4> - <\theta^2>^2) + \lambda^6 + ..., \tag{F2}$$

One sees that $\Delta = 0 \Leftrightarrow \delta = 0 \Leftrightarrow \langle \cos \lambda\theta \rangle = 1 \Leftrightarrow \langle \theta \rangle = 0 \Leftrightarrow |e_0\rangle \in \Omega_0 \cap \Omega$. In all other cases,

$$\frac{\delta^2}{\Delta^2} = \frac{\lambda^2}{4}(\frac{\langle\theta^4\rangle}{\langle\theta^2\rangle} - \langle\theta^2\rangle) + \lambda^4 ... \tag{F3}$$

Thus $\delta$ vanishes so fast for $\lambda \to 0$ that $\frac{\delta}{\Delta}$ vanishes even though $\Delta$ also does so.

For an illustration of this result consider the 3-dimensional case from sec. 3 with the specific example $|e_0\rangle = 2^{-\frac{1}{2}}(|z_0 1\rangle + |x_0 1\rangle)$. With $\lambda = 1$, $\langle z_0 1|\theta|z_0 1\rangle = 0$, $\cos(0) = 1$ and $\langle x_0 1|\theta|x_0 1\rangle = \theta_1$ one finds

$$\langle \cos\theta \rangle = \frac{1+\cos\theta_1}{2}, \quad \langle \cos^2\theta \rangle^{\frac{1}{2}} = (\frac{1+\cos^2\theta_1}{2})^{\frac{1}{2}} \text{ and } \langle \theta^n \rangle = \frac{\theta_1^n}{2}. \tag{F4}$$

Expanding $\Delta^2$ and $\delta^2$ through $\theta_1^2$ and $\theta_1^4$, the formulas (F2) give

$$\frac{\Delta^2}{2} = 1 - (\frac{1+\cos^2\theta_1}{2})^{\frac{1}{2}} = \frac{1}{4}\theta_1^2 + ... \;,$$

$$\frac{\delta^2}{2} = 1 - (\frac{1+\cos^2\theta_1}{2})^{-\frac{1}{2}}\frac{1+\cos\theta_1}{2} = \frac{1}{32}\theta_1^4 + .... \text{ and } \frac{\delta^2}{\Delta^2} = \frac{\theta_1^2}{8} + ... \tag{F5}$$

Here the exact expressions are easily checked by elementary 3-dimensional geometry (with $|e_c\rangle = 2^{-\frac{1}{2}}(|z1\rangle + |x1\rangle)$ and $P|e_0\rangle = 2^{-\frac{1}{2}}(|z1\rangle + |x1\rangle\cos\theta_1)$). The approximate expressions are checked subsequently by expansion ($\cos\theta_1 = 1 - \frac{1}{2!}\theta_1^2 + ...$). With $\theta_1$ as large as $\theta_1 = \frac{\pi}{12}$ one finds

$\Delta = 0.1838 / -0.7\%$, $\delta = 0.0173 / 1.1\%$ and $\frac{\delta}{\Delta} = 0.0943 / 1.8\%$ where the percentages indicate the errors going with use of the approximate formulas.